\documentclass[longauth]{aaEC1}\usepackage{euclid}
\usepackage{graphicx}
\usepackage{txfonts}
\usepackage{comment}
\usepackage{adjustbox}
\usepackage{tabularx}
\usepackage{lastpage}
\usepackage{csquotes}
\usepackage[switch, modulo]{lineno}

\makeatletter
\def\aa@@strip{%
    \dp\aa@stripbox \z@              
    \twocolumn[\box\aa@stripbox]     
}
\makeatother

\usepackage[breaklinks=true, colorlinks=true, citecolor=blue, linkcolor=blue, urlcolor=blue]{hyperref}

\begin{document}

\title{\Euclid: A machine-learning search for dual and lensed AGN at sub-arcsec separations\thanks{This paper is published on behalf of the Euclid Consortium.}}    

\newcommand{\orcid}[1]{} 
\author{L.~Ulivi\orcid{0009-0001-3291-5382}\thanks{\email{lorenzo.ulivi@inaf.it}}\inst{\ref{aff1},\ref{aff2},\ref{aff3}}
\and F.~Mannucci\orcid{0000-0002-4803-2381}\inst{\ref{aff3}}
\and M.~Scialpi\orcid{0009-0006-5100-4986}\inst{\ref{aff2},\ref{aff1},\ref{aff3}}
\and C.~Marconcini\orcid{0000-0002-3194-5416}\inst{\ref{aff2},\ref{aff3}}
\and G.~Cresci\orcid{0000-0002-5281-1417}\inst{\ref{aff3}}
\and A.~Marconi\orcid{0000-0002-9889-4238}\inst{\ref{aff2},\ref{aff3}}
\and A.~Feltre\orcid{0000-0001-6865-2871}\inst{\ref{aff3}}
\and M.~Ginolfi\orcid{0000-0002-9122-1700}\inst{\ref{aff2},\ref{aff3}}
\and F.~Ricci\orcid{0000-0001-5742-5980}\inst{\ref{aff4},\ref{aff5}}
\and D.~Sluse\orcid{0000-0001-6116-2095}\inst{\ref{aff6}}
\and F.~Belfiore\inst{\ref{aff3},\ref{aff7}}
\and E.~Bertola\orcid{0000-0001-5487-2830}\inst{\ref{aff3}}
\and C.~Bracci\orcid{0009-0009-0580-2604}\inst{\ref{aff2},\ref{aff3}}
\and E.~Cataldi\inst{\ref{aff3},\ref{aff2}}
\and M.~Ceci\orcid{0009-0002-2613-9564}\inst{\ref{aff2},\ref{aff3}}
\and Q.~D'Amato\orcid{0000-0002-9948-0897}\inst{\ref{aff3}}
\and I.~Lamperti\orcid{0000-0003-3336-5498}\inst{\ref{aff2},\ref{aff3}}
\and R.~B.~Metcalf\orcid{0000-0003-3167-2574}\inst{\ref{aff8},\ref{aff9}}
\and B.~Moreschini\inst{\ref{aff2},\ref{aff3}}
\and M.~Perna\orcid{0000-0002-0362-5941}\inst{\ref{aff10}}
\and G.~Tozzi\orcid{0000-0003-4226-7777}\inst{\ref{aff11}}
\and G.~Venturi\orcid{0000-0001-8349-3055}\inst{\ref{aff12},\ref{aff3}}
\and M.~V.~Zanchettin\orcid{0000-0001-7883-496X}\inst{\ref{aff3}}
\and Y.~Fu\orcid{0000-0002-0759-0504}\inst{\ref{aff13},\ref{aff14}}
\and M.~Huertas-Company\orcid{0000-0002-1416-8483}\inst{\ref{aff15},\ref{aff16},\ref{aff17},\ref{aff18}}
\and N.~E.~P.~Lines\orcid{0009-0004-7751-1914}\inst{\ref{aff19}}
\and M.~Mezcua\orcid{0000-0003-4440-259X}\inst{\ref{aff20},\ref{aff21}}
\and M.~P\"ontinen\orcid{0000-0001-5442-2530}\inst{\ref{aff22}}
\and K.~Rojas\orcid{0000-0003-1391-6854}\inst{\ref{aff23}}
\and V.~Scottez\orcid{0009-0008-3864-940X}\inst{\ref{aff24},\ref{aff25}}
\and M.~Siudek\orcid{0000-0002-2949-2155}\inst{\ref{aff16},\ref{aff20}}
\and H.~Teimoorinia\inst{\ref{aff26}}
\and I.~T.~Andika\orcid{0000-0001-6102-9526}\inst{\ref{aff27},\ref{aff28}}
\and J.~A.~Acevedo~Barroso\orcid{0000-0002-9654-1711}\inst{\ref{aff29}}
\and B.~Cl\'ement\orcid{0000-0002-7966-3661}\inst{\ref{aff29},\ref{aff30}}
\and F.~Courbin\orcid{0000-0003-0758-6510}\inst{\ref{aff31},\ref{aff32},\ref{aff33}}
\and R.~Gavazzi\orcid{0000-0002-5540-6935}\inst{\ref{aff34},\ref{aff35}}
\and L.~R.~Ecker\orcid{0009-0005-3508-2469}\inst{\ref{aff36},\ref{aff11}}
\and B.~C.~Nagam\orcid{0000-0002-3724-7694}\inst{\ref{aff37},\ref{aff14}}
\and R.~Pearce-Casey\inst{\ref{aff38}}
\and S.~Schuldt\orcid{0000-0003-2497-6334}\inst{\ref{aff39},\ref{aff40}}
\and S.~H.~Vincken\inst{\ref{aff41}}
\and D.~Stern\orcid{0000-0003-2686-9241}\inst{\ref{aff42}}
\and A.~Chakraborty\orcid{0000-0001-8016-3421}\inst{\ref{aff3}}
\and S.~Andreon\orcid{0000-0002-2041-8784}\inst{\ref{aff43}}
\and N.~Auricchio\orcid{0000-0003-4444-8651}\inst{\ref{aff9}}
\and C.~Baccigalupi\orcid{0000-0002-8211-1630}\inst{\ref{aff44},\ref{aff45},\ref{aff46},\ref{aff47}}
\and M.~Baldi\orcid{0000-0003-4145-1943}\inst{\ref{aff48},\ref{aff9},\ref{aff49}}
\and A.~Balestra\orcid{0000-0002-6967-261X}\inst{\ref{aff50}}
\and S.~Bardelli\orcid{0000-0002-8900-0298}\inst{\ref{aff9}}
\and A.~Biviano\orcid{0000-0002-0857-0732}\inst{\ref{aff45},\ref{aff44}}
\and E.~Branchini\orcid{0000-0002-0808-6908}\inst{\ref{aff51},\ref{aff52},\ref{aff43}}
\and M.~Brescia\orcid{0000-0001-9506-5680}\inst{\ref{aff53},\ref{aff54}}
\and S.~Camera\orcid{0000-0003-3399-3574}\inst{\ref{aff55},\ref{aff56},\ref{aff57}}
\and G.~Ca\~nas-Herrera\orcid{0000-0003-2796-2149}\inst{\ref{aff58},\ref{aff13}}
\and V.~Capobianco\orcid{0000-0002-3309-7692}\inst{\ref{aff57}}
\and C.~Carbone\orcid{0000-0003-0125-3563}\inst{\ref{aff40}}
\and J.~Carretero\orcid{0000-0002-3130-0204}\inst{\ref{aff59},\ref{aff60}}
\and M.~Castellano\orcid{0000-0001-9875-8263}\inst{\ref{aff5}}
\and G.~Castignani\orcid{0000-0001-6831-0687}\inst{\ref{aff9}}
\and S.~Cavuoti\orcid{0000-0002-3787-4196}\inst{\ref{aff54},\ref{aff61}}
\and A.~Cimatti\inst{\ref{aff62}}
\and C.~Colodro-Conde\inst{\ref{aff15}}
\and G.~Congedo\orcid{0000-0003-2508-0046}\inst{\ref{aff63}}
\and C.~J.~Conselice\orcid{0000-0003-1949-7638}\inst{\ref{aff64}}
\and L.~Conversi\orcid{0000-0002-6710-8476}\inst{\ref{aff65},\ref{aff66}}
\and Y.~Copin\orcid{0000-0002-5317-7518}\inst{\ref{aff67}}
\and H.~M.~Courtois\orcid{0000-0003-0509-1776}\inst{\ref{aff68}}
\and M.~Cropper\orcid{0000-0003-4571-9468}\inst{\ref{aff69}}
\and A.~Da~Silva\orcid{0000-0002-6385-1609}\inst{\ref{aff70},\ref{aff71}}
\and H.~Degaudenzi\orcid{0000-0002-5887-6799}\inst{\ref{aff72}}
\and G.~De~Lucia\orcid{0000-0002-6220-9104}\inst{\ref{aff45}}
\and A.~M.~Di~Giorgio\orcid{0000-0002-4767-2360}\inst{\ref{aff73}}
\and C.~Dolding\orcid{0009-0003-7199-6108}\inst{\ref{aff69}}
\and H.~Dole\orcid{0000-0002-9767-3839}\inst{\ref{aff74}}
\and F.~Dubath\orcid{0000-0002-6533-2810}\inst{\ref{aff72}}
\and C.~A.~J.~Duncan\orcid{0009-0003-3573-0791}\inst{\ref{aff63}}
\and X.~Dupac\inst{\ref{aff66}}
\and S.~Dusini\orcid{0000-0002-1128-0664}\inst{\ref{aff75}}
\and S.~Escoffier\orcid{0000-0002-2847-7498}\inst{\ref{aff76}}
\and M.~Fabricius\orcid{0000-0002-7025-6058}\inst{\ref{aff11},\ref{aff36}}
\and M.~Farina\orcid{0000-0002-3089-7846}\inst{\ref{aff73}}
\and R.~Farinelli\inst{\ref{aff9}}
\and F.~Faustini\orcid{0000-0001-6274-5145}\inst{\ref{aff5},\ref{aff77}}
\and S.~Ferriol\inst{\ref{aff67}}
\and F.~Finelli\orcid{0000-0002-6694-3269}\inst{\ref{aff9},\ref{aff78}}
\and M.~Frailis\orcid{0000-0002-7400-2135}\inst{\ref{aff45}}
\and E.~Franceschi\orcid{0000-0002-0585-6591}\inst{\ref{aff9}}
\and M.~Fumana\orcid{0000-0001-6787-5950}\inst{\ref{aff40}}
\and S.~Galeotta\orcid{0000-0002-3748-5115}\inst{\ref{aff45}}
\and K.~George\orcid{0000-0002-1734-8455}\inst{\ref{aff79}}
\and B.~Gillis\orcid{0000-0002-4478-1270}\inst{\ref{aff63}}
\and C.~Giocoli\orcid{0000-0002-9590-7961}\inst{\ref{aff9},\ref{aff49}}
\and J.~Gracia-Carpio\inst{\ref{aff11}}
\and A.~Grazian\orcid{0000-0002-5688-0663}\inst{\ref{aff50}}
\and F.~Grupp\inst{\ref{aff11},\ref{aff36}}
\and S.~V.~H.~Haugan\orcid{0000-0001-9648-7260}\inst{\ref{aff80}}
\and W.~Holmes\inst{\ref{aff42}}
\and I.~M.~Hook\orcid{0000-0002-2960-978X}\inst{\ref{aff81}}
\and F.~Hormuth\inst{\ref{aff82}}
\and A.~Hornstrup\orcid{0000-0002-3363-0936}\inst{\ref{aff83},\ref{aff84}}
\and K.~Jahnke\orcid{0000-0003-3804-2137}\inst{\ref{aff85}}
\and M.~Jhabvala\inst{\ref{aff86}}
\and B.~Joachimi\orcid{0000-0001-7494-1303}\inst{\ref{aff87}}
\and E.~Keih\"anen\orcid{0000-0003-1804-7715}\inst{\ref{aff88}}
\and S.~Kermiche\orcid{0000-0002-0302-5735}\inst{\ref{aff76}}
\and A.~Kiessling\orcid{0000-0002-2590-1273}\inst{\ref{aff42}}
\and B.~Kubik\orcid{0009-0006-5823-4880}\inst{\ref{aff67}}
\and M.~K\"ummel\orcid{0000-0003-2791-2117}\inst{\ref{aff36}}
\and M.~Kunz\orcid{0000-0002-3052-7394}\inst{\ref{aff89}}
\and H.~Kurki-Suonio\orcid{0000-0002-4618-3063}\inst{\ref{aff22},\ref{aff90}}
\and A.~M.~C.~Le~Brun\orcid{0000-0002-0936-4594}\inst{\ref{aff91}}
\and S.~Ligori\orcid{0000-0003-4172-4606}\inst{\ref{aff57}}
\and P.~B.~Lilje\orcid{0000-0003-4324-7794}\inst{\ref{aff80}}
\and V.~Lindholm\orcid{0000-0003-2317-5471}\inst{\ref{aff22},\ref{aff90}}
\and I.~Lloro\orcid{0000-0001-5966-1434}\inst{\ref{aff92}}
\and G.~Mainetti\orcid{0000-0003-2384-2377}\inst{\ref{aff93}}
\and D.~Maino\inst{\ref{aff39},\ref{aff40},\ref{aff94}}
\and E.~Maiorano\orcid{0000-0003-2593-4355}\inst{\ref{aff9}}
\and O.~Mansutti\orcid{0000-0001-5758-4658}\inst{\ref{aff45}}
\and O.~Marggraf\orcid{0000-0001-7242-3852}\inst{\ref{aff95}}
\and M.~Martinelli\orcid{0000-0002-6943-7732}\inst{\ref{aff5},\ref{aff96}}
\and N.~Martinet\orcid{0000-0003-2786-7790}\inst{\ref{aff34}}
\and F.~Marulli\orcid{0000-0002-8850-0303}\inst{\ref{aff8},\ref{aff9},\ref{aff49}}
\and R.~J.~Massey\orcid{0000-0002-6085-3780}\inst{\ref{aff97}}
\and E.~Medinaceli\orcid{0000-0002-4040-7783}\inst{\ref{aff9}}
\and S.~Mei\orcid{0000-0002-2849-559X}\inst{\ref{aff98},\ref{aff99}}
\and M.~Melchior\inst{\ref{aff41}}
\and Y.~Mellier\inst{\ref{aff24},\ref{aff35}}
\and M.~Meneghetti\orcid{0000-0003-1225-7084}\inst{\ref{aff9},\ref{aff49}}
\and E.~Merlin\orcid{0000-0001-6870-8900}\inst{\ref{aff5}}
\and G.~Meylan\inst{\ref{aff29}}
\and A.~Mora\orcid{0000-0002-1922-8529}\inst{\ref{aff100}}
\and M.~Moresco\orcid{0000-0002-7616-7136}\inst{\ref{aff8},\ref{aff9}}
\and L.~Moscardini\orcid{0000-0002-3473-6716}\inst{\ref{aff8},\ref{aff9},\ref{aff49}}
\and R.~Nakajima\orcid{0009-0009-1213-7040}\inst{\ref{aff95}}
\and C.~Neissner\orcid{0000-0001-8524-4968}\inst{\ref{aff101},\ref{aff60}}
\and R.~C.~Nichol\orcid{0000-0003-0939-6518}\inst{\ref{aff102}}
\and S.-M.~Niemi\orcid{0009-0005-0247-0086}\inst{\ref{aff58}}
\and C.~Padilla\orcid{0000-0001-7951-0166}\inst{\ref{aff101}}
\and S.~Paltani\orcid{0000-0002-8108-9179}\inst{\ref{aff72}}
\and F.~Pasian\orcid{0000-0002-4869-3227}\inst{\ref{aff45}}
\and K.~Pedersen\inst{\ref{aff103}}
\and W.~J.~Percival\orcid{0000-0002-0644-5727}\inst{\ref{aff104},\ref{aff105},\ref{aff106}}
\and V.~Pettorino\orcid{0000-0002-4203-9320}\inst{\ref{aff58}}
\and S.~Pires\orcid{0000-0002-0249-2104}\inst{\ref{aff107}}
\and G.~Polenta\orcid{0000-0003-4067-9196}\inst{\ref{aff77}}
\and M.~Poncet\inst{\ref{aff108}}
\and L.~A.~Popa\inst{\ref{aff109}}
\and L.~Pozzetti\orcid{0000-0001-7085-0412}\inst{\ref{aff9}}
\and F.~Raison\orcid{0000-0002-7819-6918}\inst{\ref{aff11}}
\and A.~Renzi\orcid{0000-0001-9856-1970}\inst{\ref{aff110},\ref{aff75}}
\and J.~Rhodes\orcid{0000-0002-4485-8549}\inst{\ref{aff42}}
\and G.~Riccio\inst{\ref{aff54}}
\and E.~Romelli\orcid{0000-0003-3069-9222}\inst{\ref{aff45}}
\and M.~Roncarelli\orcid{0000-0001-9587-7822}\inst{\ref{aff9}}
\and R.~Saglia\orcid{0000-0003-0378-7032}\inst{\ref{aff36},\ref{aff11}}
\and Z.~Sakr\orcid{0000-0002-4823-3757}\inst{\ref{aff111},\ref{aff112},\ref{aff113}}
\and D.~Sapone\orcid{0000-0001-7089-4503}\inst{\ref{aff114}}
\and B.~Sartoris\orcid{0000-0003-1337-5269}\inst{\ref{aff36},\ref{aff45}}
\and M.~Schirmer\orcid{0000-0003-2568-9994}\inst{\ref{aff85}}
\and P.~Schneider\orcid{0000-0001-8561-2679}\inst{\ref{aff95}}
\and T.~Schrabback\orcid{0000-0002-6987-7834}\inst{\ref{aff115}}
\and A.~Secroun\orcid{0000-0003-0505-3710}\inst{\ref{aff76}}
\and G.~Seidel\orcid{0000-0003-2907-353X}\inst{\ref{aff85}}
\and S.~Serrano\orcid{0000-0002-0211-2861}\inst{\ref{aff21},\ref{aff116},\ref{aff20}}
\and P.~Simon\inst{\ref{aff95}}
\and C.~Sirignano\orcid{0000-0002-0995-7146}\inst{\ref{aff110},\ref{aff75}}
\and G.~Sirri\orcid{0000-0003-2626-2853}\inst{\ref{aff49}}
\and L.~Stanco\orcid{0000-0002-9706-5104}\inst{\ref{aff75}}
\and J.~Steinwagner\orcid{0000-0001-7443-1047}\inst{\ref{aff11}}
\and P.~Tallada-Cresp\'{i}\orcid{0000-0002-1336-8328}\inst{\ref{aff59},\ref{aff60}}
\and D.~Tavagnacco\orcid{0000-0001-7475-9894}\inst{\ref{aff45}}
\and A.~N.~Taylor\inst{\ref{aff63}}
\and I.~Tereno\orcid{0000-0002-4537-6218}\inst{\ref{aff70},\ref{aff117}}
\and N.~Tessore\orcid{0000-0002-9696-7931}\inst{\ref{aff87},\ref{aff69}}
\and S.~Toft\orcid{0000-0003-3631-7176}\inst{\ref{aff118},\ref{aff119}}
\and R.~Toledo-Moreo\orcid{0000-0002-2997-4859}\inst{\ref{aff120}}
\and F.~Torradeflot\orcid{0000-0003-1160-1517}\inst{\ref{aff60},\ref{aff59}}
\and I.~Tutusaus\orcid{0000-0002-3199-0399}\inst{\ref{aff112}}
\and E.~A.~Valentijn\inst{\ref{aff14}}
\and L.~Valenziano\orcid{0000-0002-1170-0104}\inst{\ref{aff9},\ref{aff78}}
\and J.~Valiviita\orcid{0000-0001-6225-3693}\inst{\ref{aff22},\ref{aff90}}
\and T.~Vassallo\orcid{0000-0001-6512-6358}\inst{\ref{aff45},\ref{aff79}}
\and G.~Verdoes~Kleijn\orcid{0000-0001-5803-2580}\inst{\ref{aff14}}
\and A.~Veropalumbo\orcid{0000-0003-2387-1194}\inst{\ref{aff43},\ref{aff52},\ref{aff51}}
\and Y.~Wang\orcid{0000-0002-4749-2984}\inst{\ref{aff121}}
\and J.~Weller\orcid{0000-0002-8282-2010}\inst{\ref{aff36},\ref{aff11}}
\and G.~Zamorani\orcid{0000-0002-2318-301X}\inst{\ref{aff9}}
\and F.~M.~Zerbi\inst{\ref{aff43}}
\and E.~Zucca\orcid{0000-0002-5845-8132}\inst{\ref{aff9}}
\and A.~A.~Nucita\inst{\ref{aff122},\ref{aff123},\ref{aff124}}}
										   
\institute{University of Trento, Via Sommarive 14, I-38123 Trento, Italy\label{aff1}
\and
Dipartimento di Fisica e Astronomia, Universit\`{a} di Firenze, via G. Sansone 1, 50019 Sesto Fiorentino, Firenze, Italy\label{aff2}
\and
INAF-Osservatorio Astrofisico di Arcetri, Largo E. Fermi 5, 50125, Firenze, Italy\label{aff3}
\and
Department of Mathematics and Physics, Roma Tre University, Via della Vasca Navale 84, 00146 Rome, Italy\label{aff4}
\and
INAF-Osservatorio Astronomico di Roma, Via Frascati 33, 00078 Monteporzio Catone, Italy\label{aff5}
\and
STAR Institute, University of Li{\`e}ge, Quartier Agora, All\'ee du six Ao\^ut 19c, 4000 Li\`ege, Belgium\label{aff6}
\and
European Southern Observatory, Karl-Schwarzschild-Str.~2, 85748 Garching, Germany\label{aff7}
\and
Dipartimento di Fisica e Astronomia "Augusto Righi" - Alma Mater Studiorum Universit\`a di Bologna, via Piero Gobetti 93/2, 40129 Bologna, Italy\label{aff8}
\and
INAF-Osservatorio di Astrofisica e Scienza dello Spazio di Bologna, Via Piero Gobetti 93/3, 40129 Bologna, Italy\label{aff9}
\and
Centro de Astrobiolog\'ia (CAB), CSIC--INTA, Cra. de Ajalvir Km.~4, 28850 -- Torrej\'on de Ardoz, Madrid, Spain\label{aff10}
\and
Max Planck Institute for Extraterrestrial Physics, Giessenbachstr. 1, 85748 Garching, Germany\label{aff11}
\and
Scuola Normale Superiore, Piazza dei Cavalieri 7, 56126 Pisa, Italy\label{aff12}
\and
Leiden Observatory, Leiden University, Einsteinweg 55, 2333 CC Leiden, The Netherlands\label{aff13}
\and
Kapteyn Astronomical Institute, University of Groningen, PO Box 800, 9700 AV Groningen, The Netherlands\label{aff14}
\and
Instituto de Astrof\'{\i}sica de Canarias, V\'{\i}a L\'actea, 38205 La Laguna, Tenerife, Spain\label{aff15}
\and
Instituto de Astrof\'isica de Canarias (IAC); Departamento de Astrof\'isica, Universidad de La Laguna (ULL), 38200, La Laguna, Tenerife, Spain\label{aff16}
\and
Universit\'e PSL, Observatoire de Paris, Sorbonne Universit\'e, CNRS, LERMA, 75014, Paris, France\label{aff17}
\and
Universit\'e Paris-Cit\'e, 5 Rue Thomas Mann, 75013, Paris, France\label{aff18}
\and
Institute of Cosmology and Gravitation, University of Portsmouth, Portsmouth PO1 3FX, UK\label{aff19}
\and
Institute of Space Sciences (ICE, CSIC), Campus UAB, Carrer de Can Magrans, s/n, 08193 Barcelona, Spain\label{aff20}
\and
Institut d'Estudis Espacials de Catalunya (IEEC),  Edifici RDIT, Campus UPC, 08860 Castelldefels, Barcelona, Spain\label{aff21}
\and
Department of Physics, P.O. Box 64, University of Helsinki, 00014 Helsinki, Finland\label{aff22}
\and
University of Applied Sciences and Arts of Northwestern Switzerland, School of Computer Science, 5210 Windisch, Switzerland\label{aff23}
\and
Institut d'Astrophysique de Paris, 98bis Boulevard Arago, 75014, Paris, France\label{aff24}
\and
ICL, Junia, Universit\'e Catholique de Lille, LITL, 59000 Lille, France\label{aff25}
\and
Herzberg Astronomy and Astrophysics Research Centre, 5071 W. Saanich Rd. Victoria, BC, V9E 2E7, Canada\label{aff26}
\and
Technical University of Munich, TUM School of Natural Sciences, Physics Department, James-Franck-Str.~1, 85748 Garching, Germany\label{aff27}
\and
Max-Planck-Institut f\"ur Astrophysik, Karl-Schwarzschild-Str.~1, 85748 Garching, Germany\label{aff28}
\and
Institute of Physics, Laboratory of Astrophysics, Ecole Polytechnique F\'ed\'erale de Lausanne (EPFL), Observatoire de Sauverny, 1290 Versoix, Switzerland\label{aff29}
\and
SCITAS, Ecole Polytechnique F\'ed\'erale de Lausanne (EPFL), 1015 Lausanne, Switzerland\label{aff30}
\and
Institut de Ci\`{e}ncies del Cosmos (ICCUB), Universitat de Barcelona (IEEC-UB), Mart\'{i} i Franqu\`{e}s 1, 08028 Barcelona, Spain\label{aff31}
\and
Instituci\'o Catalana de Recerca i Estudis Avan\c{c}ats (ICREA), Passeig de Llu\'{\i}s Companys 23, 08010 Barcelona, Spain\label{aff32}
\and
Institut de Ciencies de l'Espai (IEEC-CSIC), Campus UAB, Carrer de Can Magrans, s/n Cerdanyola del Vall\'es, 08193 Barcelona, Spain\label{aff33}
\and
Aix-Marseille Universit\'e, CNRS, CNES, LAM, Marseille, France\label{aff34}
\and
Institut d'Astrophysique de Paris, UMR 7095, CNRS, and Sorbonne Universit\'e, 98 bis boulevard Arago, 75014 Paris, France\label{aff35}
\and
Universit\"ats-Sternwarte M\"unchen, Fakult\"at f\"ur Physik, Ludwig-Maximilians-Universit\"at M\"unchen, Scheinerstrasse 1, 81679 M\"unchen, Germany\label{aff36}
\and
Minnesota Institute for Astrophysics, University of Minnesota, 116 Church St SE, Minneapolis, MN 55455, USA\label{aff37}
\and
School of Physical Sciences, The Open University, Milton Keynes, MK7 6AA, UK\label{aff38}
\and
Dipartimento di Fisica "Aldo Pontremoli", Universit\`a degli Studi di Milano, Via Celoria 16, 20133 Milano, Italy\label{aff39}
\and
INAF-IASF Milano, Via Alfonso Corti 12, 20133 Milano, Italy\label{aff40}
\and
University of Applied Sciences and Arts of Northwestern Switzerland, School of Engineering, 5210 Windisch, Switzerland\label{aff41}
\and
Jet Propulsion Laboratory, California Institute of Technology, 4800 Oak Grove Drive, Pasadena, CA, 91109, USA\label{aff42}
\and
INAF-Osservatorio Astronomico di Brera, Via Brera 28, 20122 Milano, Italy\label{aff43}
\and
IFPU, Institute for Fundamental Physics of the Universe, via Beirut 2, 34151 Trieste, Italy\label{aff44}
\and
INAF-Osservatorio Astronomico di Trieste, Via G. B. Tiepolo 11, 34143 Trieste, Italy\label{aff45}
\and
INFN, Sezione di Trieste, Via Valerio 2, 34127 Trieste TS, Italy\label{aff46}
\and
SISSA, International School for Advanced Studies, Via Bonomea 265, 34136 Trieste TS, Italy\label{aff47}
\and
Dipartimento di Fisica e Astronomia, Universit\`a di Bologna, Via Gobetti 93/2, 40129 Bologna, Italy\label{aff48}
\and
INFN-Sezione di Bologna, Viale Berti Pichat 6/2, 40127 Bologna, Italy\label{aff49}
\and
INAF-Osservatorio Astronomico di Padova, Via dell'Osservatorio 5, 35122 Padova, Italy\label{aff50}
\and
Dipartimento di Fisica, Universit\`a di Genova, Via Dodecaneso 33, 16146, Genova, Italy\label{aff51}
\and
INFN-Sezione di Genova, Via Dodecaneso 33, 16146, Genova, Italy\label{aff52}
\and
Department of Physics "E. Pancini", University Federico II, Via Cinthia 6, 80126, Napoli, Italy\label{aff53}
\and
INAF-Osservatorio Astronomico di Capodimonte, Via Moiariello 16, 80131 Napoli, Italy\label{aff54}
\and
Dipartimento di Fisica, Universit\`a degli Studi di Torino, Via P. Giuria 1, 10125 Torino, Italy\label{aff55}
\and
INFN-Sezione di Torino, Via P. Giuria 1, 10125 Torino, Italy\label{aff56}
\and
INAF-Osservatorio Astrofisico di Torino, Via Osservatorio 20, 10025 Pino Torinese (TO), Italy\label{aff57}
\and
European Space Agency/ESTEC, Keplerlaan 1, 2201 AZ Noordwijk, The Netherlands\label{aff58}
\and
Centro de Investigaciones Energ\'eticas, Medioambientales y Tecnol\'ogicas (CIEMAT), Avenida Complutense 40, 28040 Madrid, Spain\label{aff59}
\and
Port d'Informaci\'{o} Cient\'{i}fica, Campus UAB, C. Albareda s/n, 08193 Bellaterra (Barcelona), Spain\label{aff60}
\and
INFN section of Naples, Via Cinthia 6, 80126, Napoli, Italy\label{aff61}
\and
Dipartimento di Fisica e Astronomia "Augusto Righi" - Alma Mater Studiorum Universit\`a di Bologna, Viale Berti Pichat 6/2, 40127 Bologna, Italy\label{aff62}
\and
Institute for Astronomy, University of Edinburgh, Royal Observatory, Blackford Hill, Edinburgh EH9 3HJ, UK\label{aff63}
\and
Jodrell Bank Centre for Astrophysics, Department of Physics and Astronomy, University of Manchester, Oxford Road, Manchester M13 9PL, UK\label{aff64}
\and
European Space Agency/ESRIN, Largo Galileo Galilei 1, 00044 Frascati, Roma, Italy\label{aff65}
\and
ESAC/ESA, Camino Bajo del Castillo, s/n., Urb. Villafranca del Castillo, 28692 Villanueva de la Ca\~nada, Madrid, Spain\label{aff66}
\and
Universit\'e Claude Bernard Lyon 1, CNRS/IN2P3, IP2I Lyon, UMR 5822, Villeurbanne, F-69100, France\label{aff67}
\and
UCB Lyon 1, CNRS/IN2P3, IUF, IP2I Lyon, 4 rue Enrico Fermi, 69622 Villeurbanne, France\label{aff68}
\and
Mullard Space Science Laboratory, University College London, Holmbury St Mary, Dorking, Surrey RH5 6NT, UK\label{aff69}
\and
Departamento de F\'isica, Faculdade de Ci\^encias, Universidade de Lisboa, Edif\'icio C8, Campo Grande, PT1749-016 Lisboa, Portugal\label{aff70}
\and
Instituto de Astrof\'isica e Ci\^encias do Espa\c{c}o, Faculdade de Ci\^encias, Universidade de Lisboa, Campo Grande, 1749-016 Lisboa, Portugal\label{aff71}
\and
Department of Astronomy, University of Geneva, ch. d'Ecogia 16, 1290 Versoix, Switzerland\label{aff72}
\and
INAF-Istituto di Astrofisica e Planetologia Spaziali, via del Fosso del Cavaliere, 100, 00100 Roma, Italy\label{aff73}
\and
Universit\'e Paris-Saclay, CNRS, Institut d'astrophysique spatiale, 91405, Orsay, France\label{aff74}
\and
INFN-Padova, Via Marzolo 8, 35131 Padova, Italy\label{aff75}
\and
Aix-Marseille Universit\'e, CNRS/IN2P3, CPPM, Marseille, France\label{aff76}
\and
Space Science Data Center, Italian Space Agency, via del Politecnico snc, 00133 Roma, Italy\label{aff77}
\and
INFN-Bologna, Via Irnerio 46, 40126 Bologna, Italy\label{aff78}
\and
University Observatory, LMU Faculty of Physics, Scheinerstrasse 1, 81679 Munich, Germany\label{aff79}
\and
Institute of Theoretical Astrophysics, University of Oslo, P.O. Box 1029 Blindern, 0315 Oslo, Norway\label{aff80}
\and
Department of Physics, Lancaster University, Lancaster, LA1 4YB, UK\label{aff81}
\and
Felix Hormuth Engineering, Goethestr. 17, 69181 Leimen, Germany\label{aff82}
\and
Technical University of Denmark, Elektrovej 327, 2800 Kgs. Lyngby, Denmark\label{aff83}
\and
Cosmic Dawn Center (DAWN), Denmark\label{aff84}
\and
Max-Planck-Institut f\"ur Astronomie, K\"onigstuhl 17, 69117 Heidelberg, Germany\label{aff85}
\and
NASA Goddard Space Flight Center, Greenbelt, MD 20771, USA\label{aff86}
\and
Department of Physics and Astronomy, University College London, Gower Street, London WC1E 6BT, UK\label{aff87}
\and
Department of Physics and Helsinki Institute of Physics, Gustaf H\"allstr\"omin katu 2, University of Helsinki, 00014 Helsinki, Finland\label{aff88}
\and
Universit\'e de Gen\`eve, D\'epartement de Physique Th\'eorique and Centre for Astroparticle Physics, 24 quai Ernest-Ansermet, CH-1211 Gen\`eve 4, Switzerland\label{aff89}
\and
Helsinki Institute of Physics, Gustaf H{\"a}llstr{\"o}min katu 2, University of Helsinki, 00014 Helsinki, Finland\label{aff90}
\and
Laboratoire d'etude de l'Univers et des phenomenes eXtremes, Observatoire de Paris, Universit\'e PSL, Sorbonne Universit\'e, CNRS, 92190 Meudon, France\label{aff91}
\and
SKAO, Jodrell Bank, Lower Withington, Macclesfield SK11 9FT, United Kingdom\label{aff92}
\and
Centre de Calcul de l'IN2P3/CNRS, 21 avenue Pierre de Coubertin 69627 Villeurbanne Cedex, France\label{aff93}
\and
INFN-Sezione di Milano, Via Celoria 16, 20133 Milano, Italy\label{aff94}
\and
Universit\"at Bonn, Argelander-Institut f\"ur Astronomie, Auf dem H\"ugel 71, 53121 Bonn, Germany\label{aff95}
\and
INFN-Sezione di Roma, Piazzale Aldo Moro, 2 - c/o Dipartimento di Fisica, Edificio G. Marconi, 00185 Roma, Italy\label{aff96}
\and
Department of Physics, Institute for Computational Cosmology, Durham University, South Road, Durham, DH1 3LE, UK\label{aff97}
\and
Universit\'e Paris Cit\'e, CNRS, Astroparticule et Cosmologie, 75013 Paris, France\label{aff98}
\and
CNRS-UCB International Research Laboratory, Centre Pierre Bin\'etruy, IRL2007, CPB-IN2P3, Berkeley, USA\label{aff99}
\and
Telespazio UK S.L. for European Space Agency (ESA), Camino bajo del Castillo, s/n, Urbanizacion Villafranca del Castillo, Villanueva de la Ca\~nada, 28692 Madrid, Spain\label{aff100}
\and
Institut de F\'{i}sica d'Altes Energies (IFAE), The Barcelona Institute of Science and Technology, Campus UAB, 08193 Bellaterra (Barcelona), Spain\label{aff101}
\and
School of Mathematics and Physics, University of Surrey, Guildford, Surrey, GU2 7XH, UK\label{aff102}
\and
DARK, Niels Bohr Institute, University of Copenhagen, Jagtvej 155, 2200 Copenhagen, Denmark\label{aff103}
\and
Waterloo Centre for Astrophysics, University of Waterloo, Waterloo, Ontario N2L 3G1, Canada\label{aff104}
\and
Department of Physics and Astronomy, University of Waterloo, Waterloo, Ontario N2L 3G1, Canada\label{aff105}
\and
Perimeter Institute for Theoretical Physics, Waterloo, Ontario N2L 2Y5, Canada\label{aff106}
\and
Universit\'e Paris-Saclay, Universit\'e Paris Cit\'e, CEA, CNRS, AIM, 91191, Gif-sur-Yvette, France\label{aff107}
\and
Centre National d'Etudes Spatiales -- Centre spatial de Toulouse, 18 avenue Edouard Belin, 31401 Toulouse Cedex 9, France\label{aff108}
\and
Institute of Space Science, Str. Atomistilor, nr. 409 M\u{a}gurele, Ilfov, 077125, Romania\label{aff109}
\and
Dipartimento di Fisica e Astronomia "G. Galilei", Universit\`a di Padova, Via Marzolo 8, 35131 Padova, Italy\label{aff110}
\and
Institut f\"ur Theoretische Physik, University of Heidelberg, Philosophenweg 16, 69120 Heidelberg, Germany\label{aff111}
\and
Institut de Recherche en Astrophysique et Plan\'etologie (IRAP), Universit\'e de Toulouse, CNRS, UPS, CNES, 14 Av. Edouard Belin, 31400 Toulouse, France\label{aff112}
\and
Universit\'e St Joseph; Faculty of Sciences, Beirut, Lebanon\label{aff113}
\and
Departamento de F\'isica, FCFM, Universidad de Chile, Blanco Encalada 2008, Santiago, Chile\label{aff114}
\and
Universit\"at Innsbruck, Institut f\"ur Astro- und Teilchenphysik, Technikerstr. 25/8, 6020 Innsbruck, Austria\label{aff115}
\and
Satlantis, University Science Park, Sede Bld 48940, Leioa-Bilbao, Spain\label{aff116}
\and
Instituto de Astrof\'isica e Ci\^encias do Espa\c{c}o, Faculdade de Ci\^encias, Universidade de Lisboa, Tapada da Ajuda, 1349-018 Lisboa, Portugal\label{aff117}
\and
Cosmic Dawn Center (DAWN)\label{aff118}
\and
Niels Bohr Institute, University of Copenhagen, Jagtvej 128, 2200 Copenhagen, Denmark\label{aff119}
\and
Universidad Polit\'ecnica de Cartagena, Departamento de Electr\'onica y Tecnolog\'ia de Computadoras,  Plaza del Hospital 1, 30202 Cartagena, Spain\label{aff120}
\and
Infrared Processing and Analysis Center, California Institute of Technology, Pasadena, CA 91125, USA\label{aff121}
\and
Department of Mathematics and Physics E. De Giorgi, University of Salento, Via per Arnesano, CP-I93, 73100, Lecce, Italy\label{aff122}
\and
INFN, Sezione di Lecce, Via per Arnesano, CP-193, 73100, Lecce, Italy\label{aff123}
\and
INAF-Sezione di Lecce, c/o Dipartimento Matematica e Fisica, Via per Arnesano, 73100, Lecce, Italy\label{aff124}}         

    \titlerunning{A machine-learning search for dual and lensed AGN}
    \authorrunning{Ulivi, L., et al.}

   \date{Received July 16, 2025; accepted ----}

 
  \abstract
   {Cosmological models of hierarchical structure formation predict the existence of a widespread population of dual accreting supermassive black holes (SMBHs) on kiloparsec-scale separations, corresponding to projected distances $<$ \ang{;;0.8} at redshifts higher than 0.5. However, close companions to known active galactic nuclei (AGN) or quasars (QSOs) can also be multiple images of the object itself, strongly lensed by a foreground galaxy, as well as foreground stars in a chance superposition. Thanks to its large sky coverage, sensitivity, and high spatial resolution, \Euclid  offers a unique opportunity to obtain a large, homogeneous sample of dual/lensed AGN candidates with sub-arcsec projected separations. 
   Here we present a machine learning approach, in particular a Convolutional Neural Network (CNN), to identify close companions to known QSOs down to separations of $\sim\,$\ang{;;0.15}, comparable to the \Euclid VIS point spread function (PSF). We investigated the performance of the CNN trained on a large sample of simulated single and dual AGN.    We studied the effectiveness of the CNN in identifying dual AGN  and demonstrated that it outperforms traditional techniques such as \texttt{IRAFStarFinder}. Applying our CNN to a sample of $\sim\,$6000 QSOs from the Q1 \Euclid  data release (covering \SI{63.1}{\deg\squared}), we find a fraction of about 0.25\% dual AGN candidates with separation  $\sim\,$~\ang{;;0.4} (corresponding to $\sim\,$3\,kpc at $z\,$=1). Estimating the foreground contamination from stellar objects, we find that most of the pair candidates with separation higher than \ang{;;0.5} are likely contaminants, while below this limit, contamination is expected to be less than 20\%.  For objects at higher separation (>\,~\ang{;;0.5}, i.e. 4\,kpc at $z\,$=1), we performed PSF subtraction and used colour-colour diagrams to constrain their nature. We present a first set of dual/lensed AGN candidates detected in the Q1 \Euclid data, providing a starting point for the analysis of future data releases.

   
   }


    \maketitle

\section{Introduction}
The formation and growth of supermassive black holes (SMBHs) are closely linked to the evolution of their host galaxies. In particular, galaxy mergers are thought to play a central role in driving gas inflows toward the central regions, potentially triggering both star formation and active galactic nuclei (AGN) activity \citep[e.g.][]{Hernquist1989, Hopkins2008, Blecha2017}. When two galaxies merge, the SMBHs in their nuclei become part of the same post-merger galaxy, and eventually merge together, typically on timescales of the order of Gyr \citep[e.g.][]{Khan2012}. In the final stages of a merger, both SMBHs may be simultaneously active, forming a dual AGN system (i.e. two accreting SMBHs within a single merged galaxy). These systems provide a unique laboratory for studying SMBH growth, AGN triggering, and the interplay between galaxy evolution processes and nuclear activity.

Simulations predict that dual AGN can form in both major and minor galaxy mergers \citep[e.g.,][]{Capelo2015, Capelo2017, Chen2023}, with longer observable dual-AGN lifetimes expected in major mergers. Conversely, the highest SMBH mass ratios are predicted to arise in minor mergers, which may dominate the observed population at lower luminosities ($<\,10^{44}\,\rm erg\,s^{-1}$), especially in the regime accessible to facilities such as \textit{Euclid}. 

While theoretical models predict that dual AGN should be relatively common, especially at separations below several kpc \citep[e.g.][]{Rosas-Guevara2019,Volonteri2022,Chen2023}, observational studies have found mixed results, with reported dual AGN fractions ranging from a few percent to about 25\% depending on selection method and sample properties \citep[e.g.][]{Koss2012, Perna2025}. However, these estimates are possibly affected by several biases, including reliance on visual classification which can miss obscured or compact systems, as well as luminosity cuts, maximum separation, and the selection of the sample considered. Quantifying the frequency and properties of dual AGN as a function of separation, redshift, and host galaxy properties (e.g. extinction, star formation, black hole mass) is essential for constraining models of SMBH formation and co-evolution with galaxies. In particular, galactic scales ranging from tens of kiloparsecs to tens of parsecs represent a key regime, corresponding to the late stages of mergers when SMBHs are on the path to forming bound binaries.

On the observational side, there is still a lack of robust statistical samples of kpc-scale dual AGN systems at $z > 1$. Identifying such systems poses significant challenges, primarily due to the stringent requirements on spatial resolution and reliable AGN diagnostics. Detecting dual AGN at kpc separations requires sub-arcsec resolution, while their intrinsic rarity demands wide-area searches. To date, ground-based wide-field imaging and spectroscopic surveys have proven to be largely inefficient in uncovering sub-arcsec dual and lensed at high redshift.

Recently, the all-sky \textit{Gaia} mission has enabled the identification of candidate dual and lensed AGN at sub-arcsec separations, extending to $z > 1$ \citep[e.g.][]{Lemon2017, Chen2022, Mannucci2022, Mannucci2023,Ciurlo2023,Glikman2023, Scialpi2024}. However, this approach is limited to optically bright quasars (QSOs) due to \textit{Gaia}’s magnitude limit of $G \sim 21$, which corresponds to $L_{\rm bol} \gtrsim 10^{46}$~erg~s$^{-1}$ at $z \gtrsim 0.5$. As a result, current \textit{Gaia}-selected samples probe only the most luminous AGN population, significantly biasing against more typical, moderate-luminosity AGN and limiting the completeness of such surveys.

Over the last ten years, astrophysics has undergone a significant shift in its approach, transitioning swiftly from dealing with relatively small datasets to operating with big data. 
Survey telescopes such as the ESA satellite \textit{Euclid} \citep{EuclidSkyOverview} generate vast amounts of data that require automated tools leveraging advancements in high-performance computing, machine learning, data science, visualization and foundation models \citep[e.g.\ ][]{Q1-SP052,AcevedoBarroso24,Q1-SP048,Q1-SP053,Pearce-Casey2025,Nagam25, Q1-SP049}.
Machine learning (ML) and, in particular, deep learning (DL) offer powerful data-driven approaches for an efficient analysis of large and complex datasets.

In this work, we explore the application of ML techniques to the identification of dual AGN systems in imaging data, with a particular focus on the first \Euclid quick release, Q1 \citep{Q1cite}. As dual AGN are rare and often difficult to identify, especially at high redshift and small angular separations, advanced classification methods, such as convolutional neural networks (CNNs), offer a promising approach to improve both the completeness and reliability of detections.



The present paper is structured as follows. In Sect.~\ref{dataandmethods} we describe the input catalogue of AGN, the data used in this work, which consists of the \Euclid Q1 data, and the ML training set (Sect.~\ref{simulations}) as well as the architecture of the CNN. In Sect.~\ref{sec:simulationsresults} we present the results obtained from simulations, compare our method with other standard techniques and discuss the identification of an optimal classification threshold. In Sect.~\ref{sec:applicationtoq1data} we apply our CNN to the Q1 data, we discuss foreground contamination, we extract the photometry exploiting the four \Euclid bands and present a sample of dual AGN candidates. In Sect.~\ref{sec:discussion} we compare our results with both simulation and observational works from the literature. 
Throughout this work, we assume $ \Omega_{\rm m}=0.286$ and $H_0=69.9$ \kms Mpc$^{-1}$ \citep{Bennett2014}.



\section{Data and methods}\label{dataandmethods}


\subsection{Euclid Data}\label{Eucliddata}

This work is based on imaging data from the \Euclid mission, which is conducting two major surveys: the \Euclid Wide Survey (EWS) and the \Euclid Deep Survey (EDS). The EWS \citep{Scaramella-EP1} is designed to cover approximately one-third of the sky, amounting to $\sim\,14\,000~\mathrm{deg}^2$, while avoiding the Galactic and ecliptic planes. The Deep Survey, in contrast, focuses on three carefully selected regions -- the Euclid Deep Field (EDF) North, South, and Fornax -- and achieves significantly greater depth through repeated exposures. In this study, we use data from the \Euclid Q1 release \citep{Q1-TP001}, which includes $\sim\,$63~deg$^2$ of imaging of the EDFs acquired to the depth of the Wide Survey.
We use the \texttt{pdMerBksMosaic} data products \citep{Q1-TP004}, which include background-subtracted mosaics and associated point spread function (PSF) catalogues (\texttt{catalogue-PSF}). For this analysis, we consider all Q1 fields which span approximately 350 tiles, each with a field of view of 0.57 $\mathrm{deg}^2$ , corresponding to \ang{0.75} $\times$ ~\ang{0.75}. These observations reach a $\sim$~10$\sigma$ magnitude limit of 24.5 in the Visible Instrument
(VIS) filter, \IE \citep{EuclidSkyVIS}, and a 5\,$\sigma$ magnitude limit of 24.5 in each of the filters of the Near
Infrared Spectrometer and Photometer (NISP), namely \YE, \JE, and \HE \citep{EuclidSkyNISP}.  
The Q1 data give us a first glimpse into the possibilities of efficiently searching for dual AGN at sub-arcsec separations over the full \textit{Euclid} survey of $\sim\,$14\,000 $\mathrm{deg}^2$.

\subsection{Input QSO catalogue}

\begin{table*}[]
    \centering
    \caption{List of QSO catalogues used to compile the search list.}

    \begin{tabular}{llrl}
    \hline
    \hline
    Catalogue           &  Properties  &  Number & Ref \\
    \hline
    Milliquas v8.0 & Spectroscopy from literature  &  729\,519& \cite{Flesch2023}\\
    Quaia G20.5    & \textit{Gaia} and \textit{WISE} colours          &  878\,145& \cite{Storey-Fisher2024}\\
    eRosita        & X-ray flux                    &  177\,971& \cite{Merloni2024}\\
    \textit{WISE}           & \textit{WISE} colours                   & 2\,275\,316& \cite{Assef2018}\\
    \textit{Gaia}/unWISE    & \textit{Gaia} and WISE colours          & 1\,228\,393& \cite{Shu2019}\\
    \hline
    TOT (excluding duplications)     & & 3\,295\,296 & \\
    \hline
    \end{tabular}
    \label{tab:QSOcatalogues}
\end{table*}


\begin{figure}[h]
    \centering
    \includegraphics[width=1\linewidth]{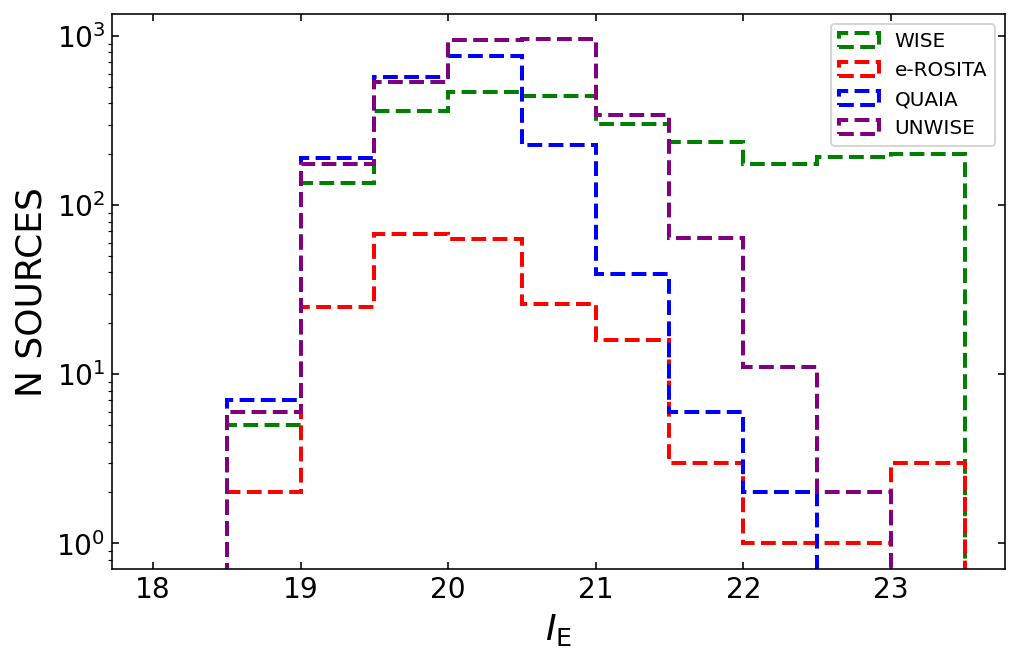}
    \caption{Number of Q1 AGN sample in $\Delta\IE=0.5$ magnitude bins as a function of the selection method: \textit{WISE} (green dashed line), e-ROSITA (red), QUAIA (blue), and UNWISE (purple).  }
    \label{fig:Distributionfunction}
\end{figure}

The dual/lensed search strategy consists of looking for close companions near previously classified QSOs. We created the input QSO list by merging several external catalogues of both spectroscopically confirmed QSOs and robust photometric QSO candidates selected in various ways. Our aim is to privilege completeness over purity, while discarding the most unreliable candidates. More accurate assessments of the nature of candidate double systems could be later performed on an individual basis, e.g. with AO-assisted spectroscopy. For each catalogue we only selected the QSOs falling within the EWS footprint, i.e. excluding objects within \ang{23} from the Galactic plane and \ang{10} from the ecliptic plane. We also excluded QSOs close to the largest galaxies in the Local Group, e.g., Large Magellanic Cloud (LMC), Small Magellanic Cloud (SMC), M31, and M33. Table~\ref{tab:QSOcatalogues} reports the  catalogues used and the number of objects obtained from each of them.
In particular:
\begin{itemize}
\item Milliquas v8.0 \citep{Flesch2023}: a compilation of all the known QSOs with a reliable redshift. 
Most of its entries are from the Sloan Digital Sky Survey (SDSS) QSO catalogue DR16Q \citep{Lyke2020}, the DESI EDR QSO catalogue \citep{Chaussidon2023}, the LAMOST survey \citep{Jin2023}, and the 2DF QSO survey \citep{Croom2004}.
\item \textit{WISE}: \cite{Assef2018} presented a catalogue of about 4.5 million QSO candidates based on \textit{WISE} colours. To decrease the number of false positives, following \cite{Stern2012}, we further select the objects with \textit{WISE} colour $W1$~$-$~$W2>$~0.8 and having $W1<$~18.5 and $W2<$~16.0 (77\% of the original sample).

\item Quaia: \cite{Storey-Fisher2024} drew an all-sky QSO catalogue combining  \textit{Gaia} candidates \citep{Gaia2016} with unWISE infrared data (based on the \textit{WISE} survey).

\item eRosita: \cite{Merloni2024} present catalogues of both point-like and extended sources using the data acquired in the first six months of survey operations (eRASS1; completed in June 2020) over  half of the sky, whose proprietary data rights lie with the German eROSITA Consortium. From their full catalogue, we selected all the sources that are clearly detected (DET\_LIKE\_0 > 9) and not extended (parameter EXT = 0).
\end{itemize}
We crossmatched the different catalogues using a $5\arcsec$ search radius to find duplicate sources. When a duplication is found, the coordinates derived from the \textit{Gaia} and spectroscopic catalogues are preferred over those from eRosita and \textit{WISE}.  The final catalogue has about 3.29 million sources (see Table~\ref{tab:QSOcatalogues}).



\subsection{Euclid VIS stamps}
\begin{figure}
    \centering
    \includegraphics[width = 0.5 \textwidth]{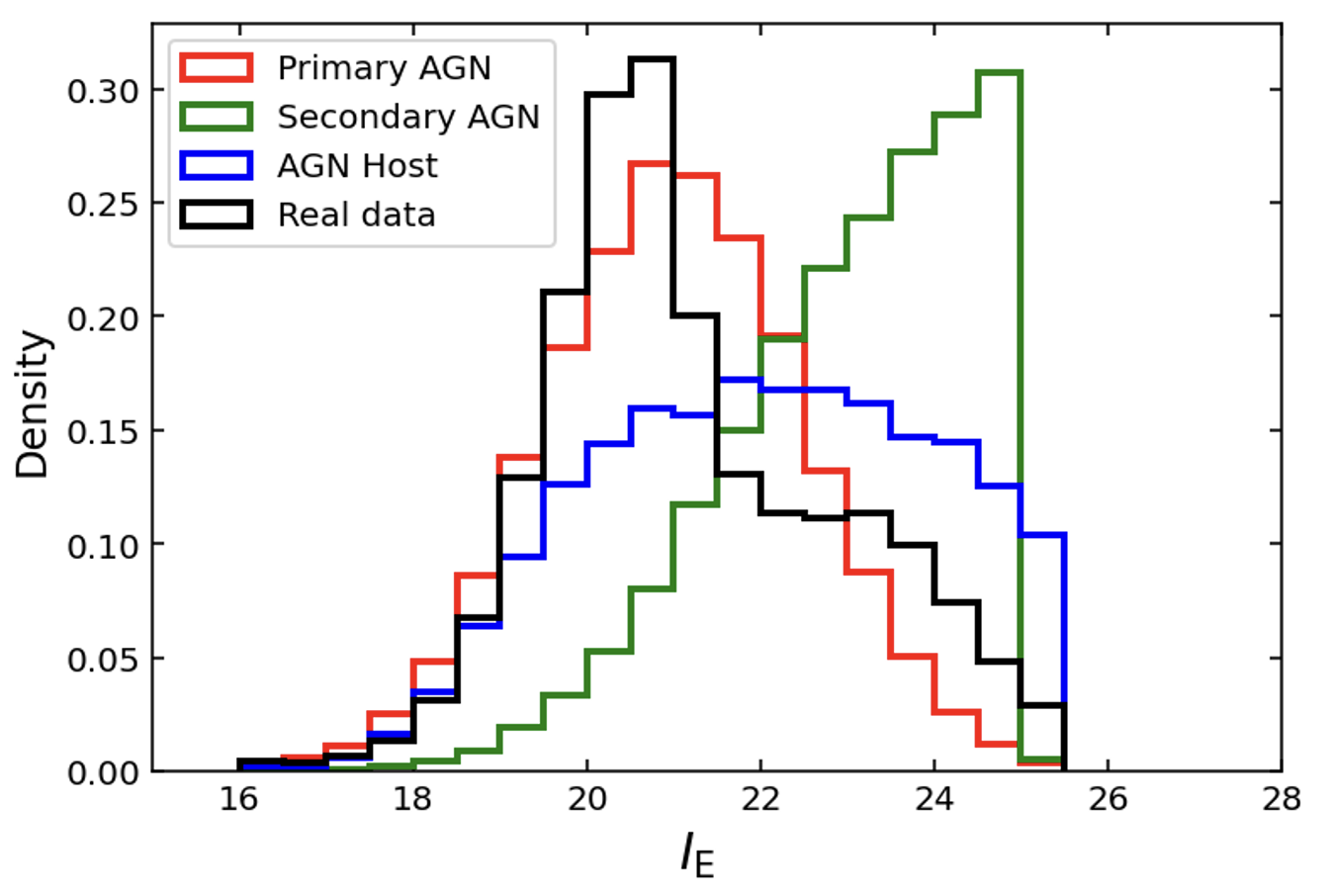}
    \caption{Distribution of \IE of the primary AGN (red), secondary AGN (green) and the associated host galaxy (blue) in simulations compared to the distribution of \IE of the considered sample of AGN in Q1 (black) in 0.5 magnitude bins.}
    \label{fig:Number Density}
\end{figure}

\begin{figure*}[t]
    \centering
    \includegraphics[width = 1 \textwidth]{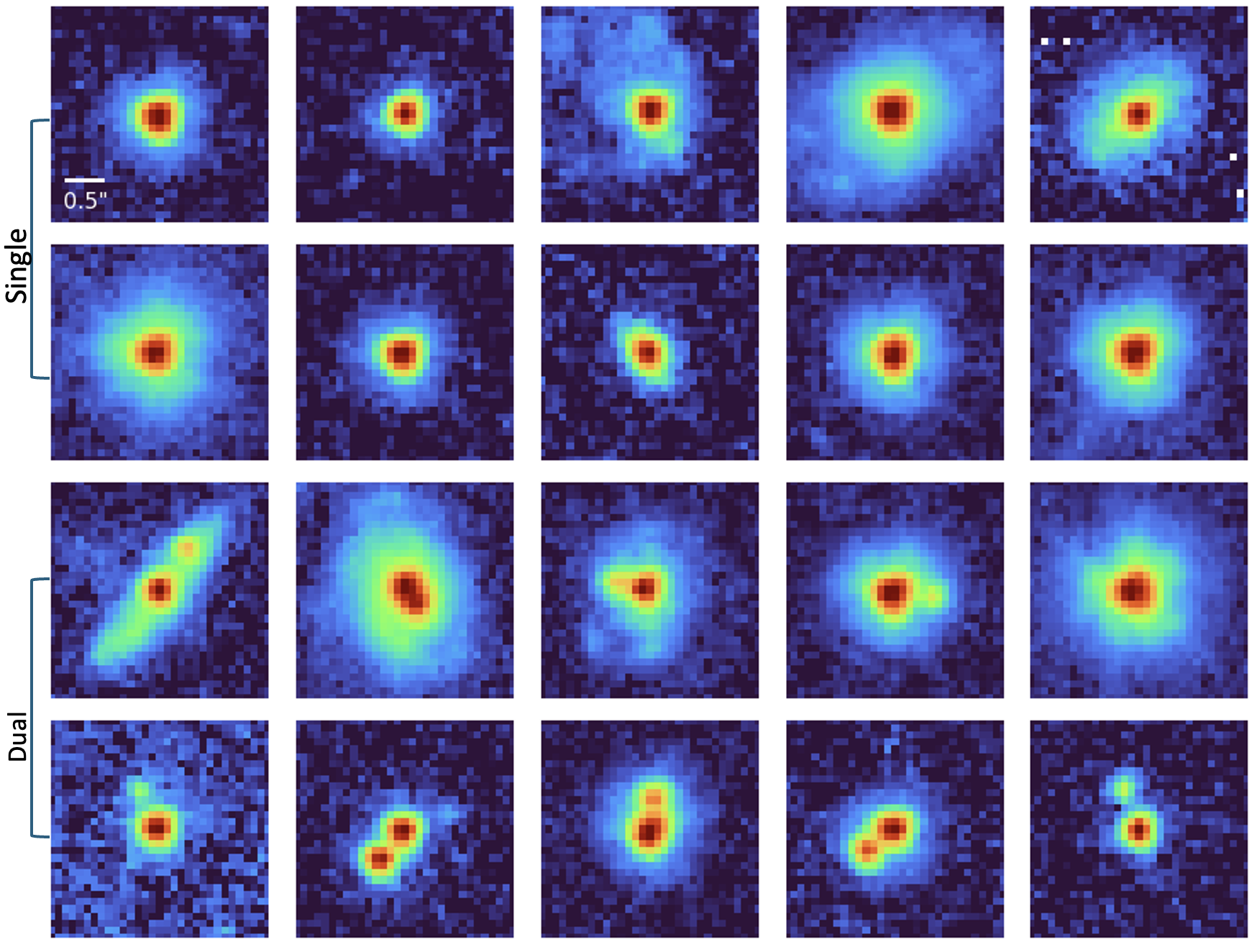}
    \caption{Examples of ten simulations of a single AGN (first two rows) and of a dual AGN (second two rows) in \IE band. The size of each cutout is $\ang{;;3} \times \ang{;;3}$. }
    \label{fig:Simulations}
\end{figure*}
Starting from the catalogue described in the previous section, we performed cutouts of the VIS images of $\ang{;;5} \times \ang{;;5}$ and 4\arcmin $\times$ 4\arcmin\ around each AGN in the catalogue, ending up with 15\,000 AGN images in the Q1 data.
We used \texttt{SExtractor} \citep{Bertin1996} for object detection in both types of cutouts, employing specific parameters to optimise performance. In particular, 
the goal to use \texttt{SExtractor} in the small ($\ang{;;5} \times \ang{;;5}$) cutouts was to measure object properties and apply selection cuts in magnitude and size for our final input sample. The larger 4\arcmin $\times$ 4\arcmin\ cutouts were processed with \texttt{SExtractor} to build a catalogue of host galaxies, which were then used in the simulation described in the following section.
The \texttt{SExtractor DETECT MINAREA} was set to 2, the DETECT THRESH was set at 1.5 times the background RMS noise, allowing us to detect faint objects. We applied a $4-\text{pixel}$ Gaussian filter (indicating that the sources are smoothed with a Gaussian kernel with $\sigma$ = 4 pixel) and set the parameter that defines the minimum contrast between the two sources to consider them separated to 0.005. These parameter choices were tailored to improve detection sensitivity. MAG ZERO is provided in the header of each tile \citep{Q1-TP004}. 
In this way, for each small cutout, we generated a source catalogue and assigned to the AGN candidate from the external catalogue the magnitude \IE of the brightest detected object within the cutout. We flagged those sources which were not centred in a square of \ang{;;0.3} around the coordinates of the external catalogue.
We centred the image in the pixel with the highest flux. 
For each AGN we keep track of the \textit{Gaia} magnitude, the \textit{WISE} colours and, the selection method. 

In Fig.~\ref{fig:Distributionfunction} we report the number of sources in 0.5 magnitude bins of \IE for different selection methods. We see that most of the QSOs in our catalogue are selected through (UN)WISE colour.



\subsection{Simulations}
\label{simulations}
The simulations described in this section were specifically designed for training and validating a neural network to detect single and dual AGN systems. Our strategy is based on adding synthetic AGN sources, represented by PSFs, into real \Euclid VIS galaxy images to create realistic examples of both single and dual AGN systems.
We generated a dataset comprising 100\,000 synthetic images, 50\,000 containing a single AGN source and 50\,000 containing a dual AGN system in the VIS band. 
Each AGN was modelled as a point source.  When downloading the Q1 tiles, we also downloaded the grid of PSFs provided by the official pipeline  (catalogue PSF). From each tile, we made cutouts of 200 randomly chosen PSFs throughout the field. 
The PSFs taken from the catalogue PSF are undersampled (19$\times$19), so we used drizzle \citep{drizzle} to oversample them using the `turbo' kernel to keep the flux constant.
We split the 100\,000 simulations (50\,000 single sources, 50\,000 double sources) into 60\% assigned to the training set and of the remaining 40\%, 20\% to the validation set and 80\% to the test set.

The magnitude of each primary AGN was drawn from a Gaussian distribution centred at \IE = 21 with a standard deviation of $\sigma$ = 1.2, chosen to closely match the observed magnitude distribution of the AGN from the external catalogue, as shown in Fig.~\ref{fig:Number Density}. 
A priori, the relative contributions of the AGN and its host galaxy to the total observed luminosity are unknown. Previous studies have shown that this contribution can vary significantly with luminosity and redshift. For instance, \citet{Rakshit2020}, analysing a sample of approximately 500\,000 QSOs at $z < 0.8$, found that the host galaxy contribution decreases with increasing total luminosity. At bolometric luminosities around $10^{45}\ \mathrm{erg\ s^{-1}}$, the host galaxy can contribute up to 80\% of the total emission. Extending to higher redshifts ($z \sim 1.5$), \citet{Schramm2013} demonstrated that, particularly for AGN with lower luminosities, the host galaxy contamination can be even more significant, even exceeding the 90\% of the total. Given this uncertainty, we developed simulations covering a broad parameter space to robustly explore the possible contributions of the AGN and its host galaxy. However, the main purpose of including host galaxies was to enhance the training of the neural network by introducing more realistic and challenging cases while preserving the underlying physics.

We associated to each point source a host galaxy with a flux between 0 and 10 times that of the point source and with CLASS\_STAR < 0.5, where CLASS\_STAR, taken from \texttt{SExtractor}, is defined as the probability to be a point source (1) or an extended object (0). In the case of host galaxies dimmer that \IE=27.5, we just added random noise to the image as they would be below the detection limit of \Euclid. To select the random noise, we computed for each tile the median and the median absolute deviation (MAD) and we made random cutouts of the FOV of the tile, accepting the cutout if the mean is less than the median and the $\sigma$ < 2 MAD. 
We designed 50\,000 simulations with single objects and we ended up with 2\% of simulated AGN painted without the host galaxy. Additionally, we added to each stamp a secondary source with a separation drawn from a uniform distribution between \ang{;;0.1} and \ang{;;0.8} from the primary AGN and whose magnitude was imposed to vary uniformly in magnitude between that of the primary and 25 mag, the limiting AB magnitudes (10\,$\sigma$ for point-like sources) achieved in each footprint. 

We explored several sets of simulations, each differing in some aspects to better capture the important features. The main differences between the simulation sets include: the distribution of magnitudes of the primary and secondary objects; the relationship between the AGN and host galaxy magnitudes; the distribution of projected separation between sources; the PSF used to model the AGN; and the level and standard deviation of the noise.  
These parameters are critical for the neural network because it has to be trained on data as close as possible to the \Euclid dataset, while learning to recognize features. The main problem of using a neural network to identify dual AGN is inherent to their rareness, which poses a serious challenge to the training of the network. For this purpose, it is necessary to simulate many of such sources, which makes the distribution of simulated objects different from that of the \Euclid data. This in itself is true for dual AGN, but it is also true for the simulated magnitude distributions of the two sources; in fact, the probability of having more faint primary and secondary objects will be larger than the probability of having bright objects, but this leads to imbalances in the simulations from which the network can learn biased features. However, in this work we focus on the simulation setup described above, which provided the best performance in training and validating our model with the Q1 data (see also Appendix~\ref{sec:comparison}). Some examples of simulated single and double AGN are shown in Fig.~\ref{fig:Simulations}.



\begin{figure*}[t]
    \centering
    \includegraphics[width = 1 \textwidth]
    {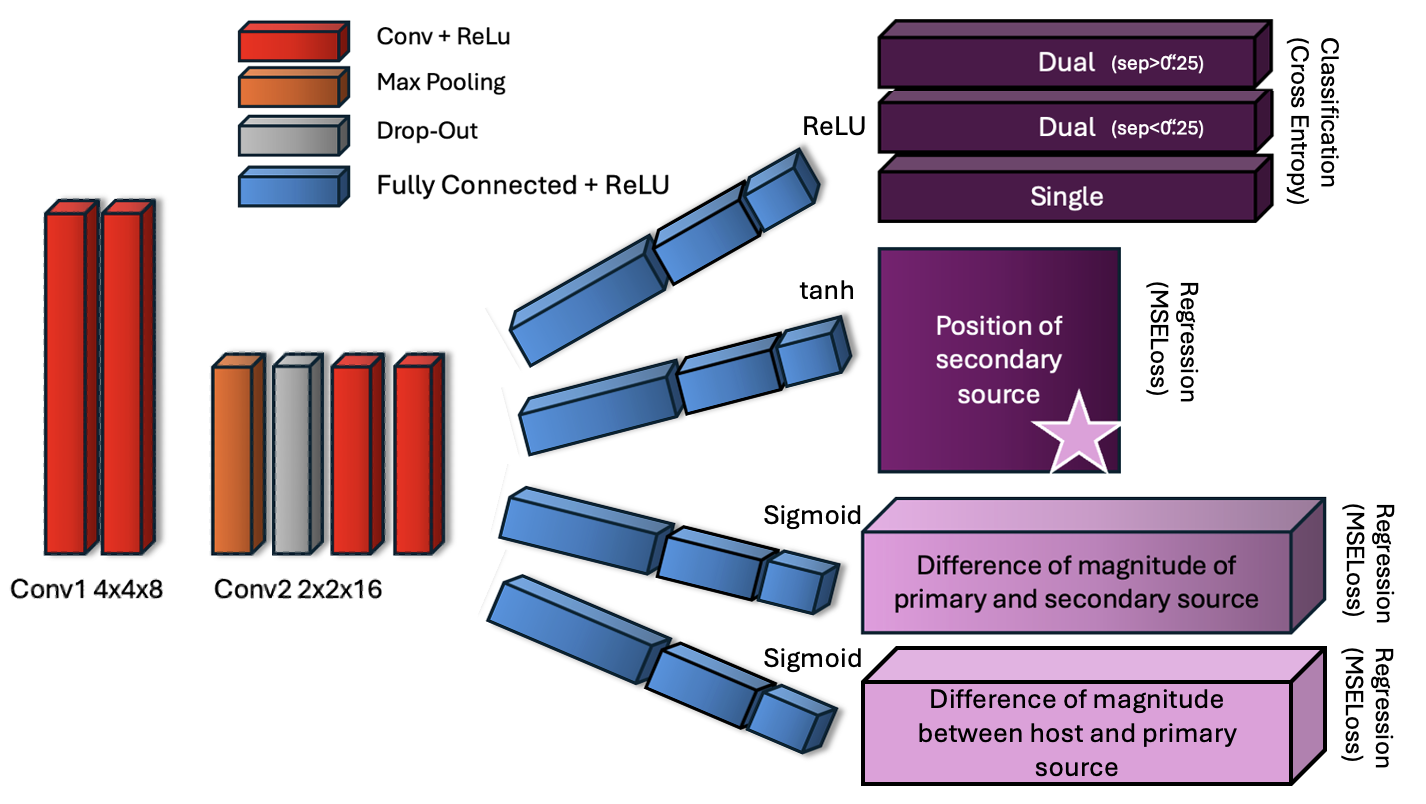}
    \caption{Architecture of the convolutional neural network used for classification and feature extraction. The model consists of two convolutional layers (Conv1: 4×4×8, Conv2: 2×2×16) with ReLU activation, followed by max pooling and dropout layers. Fully connected layers ($2304-80-20$) process the features for output into multiple targets: the position of the secondary source, the difference in magnitude between primary and secondary sources, and the difference in magnitude between the host and primary source. Each output branch uses a different final activation function and loss: ReLU and cross-entropy loss function for the classification task, hyperbolic tangent for the position, and sigmoid for the magnitude differences. The loss functions are the mean squared error when applied to regression task.}
    \label{fig:CNN}
\end{figure*}

\subsection{Convolutional neural network classifier} 
\label{Sample}
The images were labelled with 2 if they contained dual sources with separations greater than \ang{;;0.25}, 1 if they contained dual sources with separation less than \ang{;;0.25} or 0 if single sources. The \ang{;;5} $\times$ \ang{;;5} stamps are centred in the brightest pixels and cut into a 30 $\times$ 30 pixels (\ang{;;3} $\times$ \ang{;;3}), which is first processed by a 2D convolution layer with a 4 $\times$ 4 filter size, then subsampled by a 2 $\times$ 2 max pooling layer. Two more identical units follow, with a growing dimensionality of the output space in the convolution, for a total of 2 convolutional layers and 2 max pooling layers. Each of these convolutional layers is followed by a ReLU activation layer. The output of these units is then processed through three single fully-connected layers. For the classification problem we used the cross-entropy cost function and found the weights using the ADAM \citep{Kingma2014} optimisation method. Note that in PyTorch \citep{Pytorch}, no explicit softmax activation function is needed in the final layer, as it is handled internally by the cost function. The use of the ADAM optimizer improved the learning rate compared to tests with stochastic gradient descent (SGD). To prevent overfitting and enhance generalisation, we added dropout layers which randomly drop out  a fraction ($30\%$) of neurons during the training phase. 
We implemented our network as a multitask model, enabling it to predict, in addition to the nature of the system, the position of the secondary, the magnitude difference between the secondary and the primary, and the magnitude difference between the host and the primary. Multitask neural networks are powerful tools because they allow a model to learn several related tasks simultaneously, improving the performance while reducing the risk of overfitting, especially when the tasks are correlated. This approach often leads to more robust models and can capture structures in the data more effectively than training separate models for each task \citep{Ruder2017,Crawshaw2020, Ginolfi2025}.

We built another block in our CNN that predicts the coordinates $x$ and $y$ of the secondary AGN. To increase the effectiveness of the network, we made it so that the position of the secondary can vary between $-1$ and $1$, thus normalising to the maximum distance that the secondary can reach in our simulations. The activation function used for this task is a tanh function (so that it varies between $-1$ and 1). We assigned to the single sources the position of the secondary source at the value \textit{x} = 0, $\textit{y} = 0$. This can also give us an indication for detecting false positives, that is, single objects classified as dual; in fact when the network detects dual but predicts secondaries with coordinates very close to (0,0), we can discard this prediction. 

We also added two branches to the neural network to predict the magnitude difference between the primary and secondary sources ($\Delta I_{\mathrm{E,21}}$) and, though more difficult, the difference in magnitude between the primary and its host galaxy ($\Delta I_{\mathrm{E,H1}}$). For the single objects, we imposed as the true label of $\Delta I_{\mathrm{E,21}}$ the maximum among those simulated in the double objects, so as to normalise all the sample between 0 and 1 and without restricting in a smaller parameter space the double objects. We did the same for the difference between the magnitude of the primary and that of the host galaxy. We used the mean squared error as the loss function for the regression tasks. For the multi-task learning framework, we computed the total loss as the sum of all individual task losses, with the classification loss contributing more strongly to the optimization process due to its relative scale compared to the other tasks.
The final architecture of our model is illustrated in Fig.~\ref{fig:CNN}.

\section{Simulation results}\label{sec:simulationsresults}
In this section, we present the results of our neural network applied to the simulated datasets. In particular, we study how the completeness and accuracy vary as a function of both the physical parameters of the systems and the quantities predicted by the network. We also compare the performance of our approach with standard methods, and select the most suitable network architecture for the application to the real data.

\subsection{ROC curve and comparison with standard method }\label{sec:ROC}

\begin{figure}[t!]
\centering
    \includegraphics[width = 0.5 \textwidth]{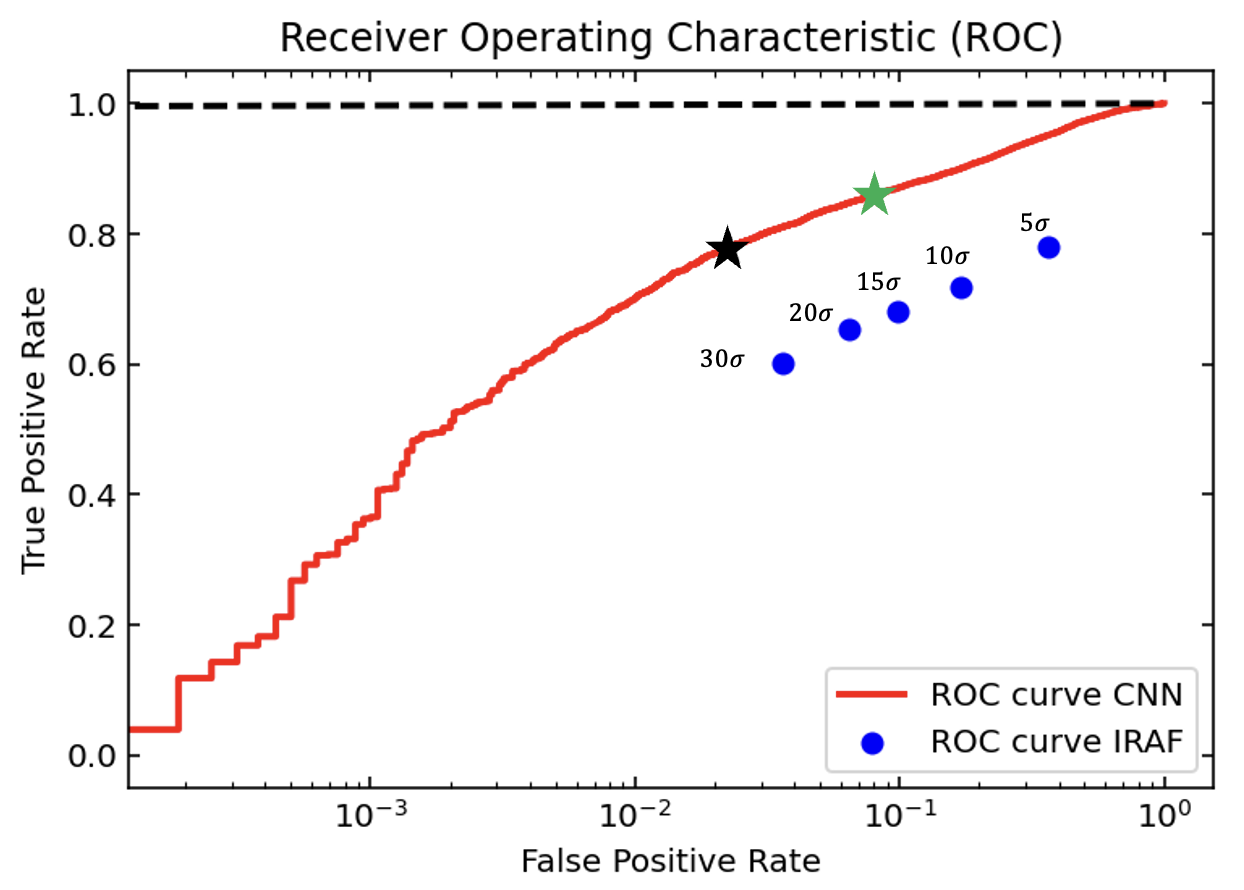}
    \caption{ROC curve comparing the performance of the two detection methods in log scale. The red curve represents the ROC of the CNN, while the blue points correspond to the ROC obtained using the \texttt{IRAFStarFinder} algorithm with varying detection thresholds (5, 10, 15, 20, 30 $\sigma$). The green star marks the threshold closest to an ideal classifier, while the black star indicates the threshold that minimises the difference between the FPs and TPs assuming a real case scenario (see Sect. \ref{sec:threshold}).  }
    \label{fig:roc_con_host}
\end{figure}

\begin{figure}
\centering
    \includegraphics[width = 0.5 \textwidth]{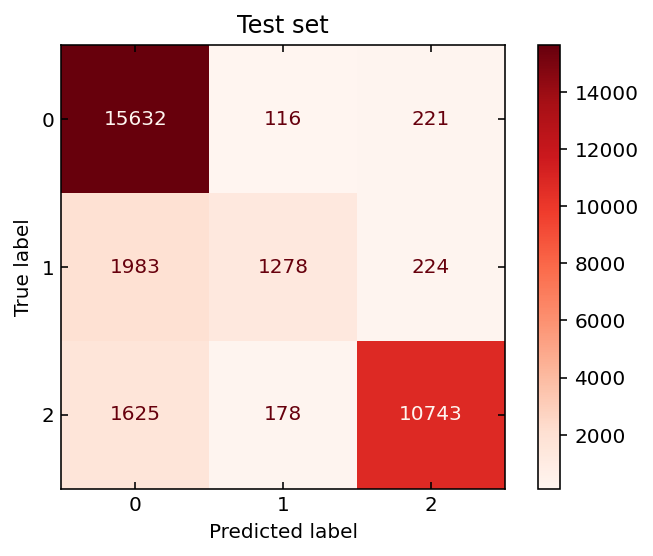}
    \caption{Confusion matrix. 0: single AGN, 1: dual AGN with separation less than \ang{;;0.25}, 2: dual AGN with separation greater than \ang{;;0.25}. }
    \label{fig:conf_matrix}
\end{figure}

\begin{figure*}[t]
    \centering
    \includegraphics[width = 1 \textwidth]{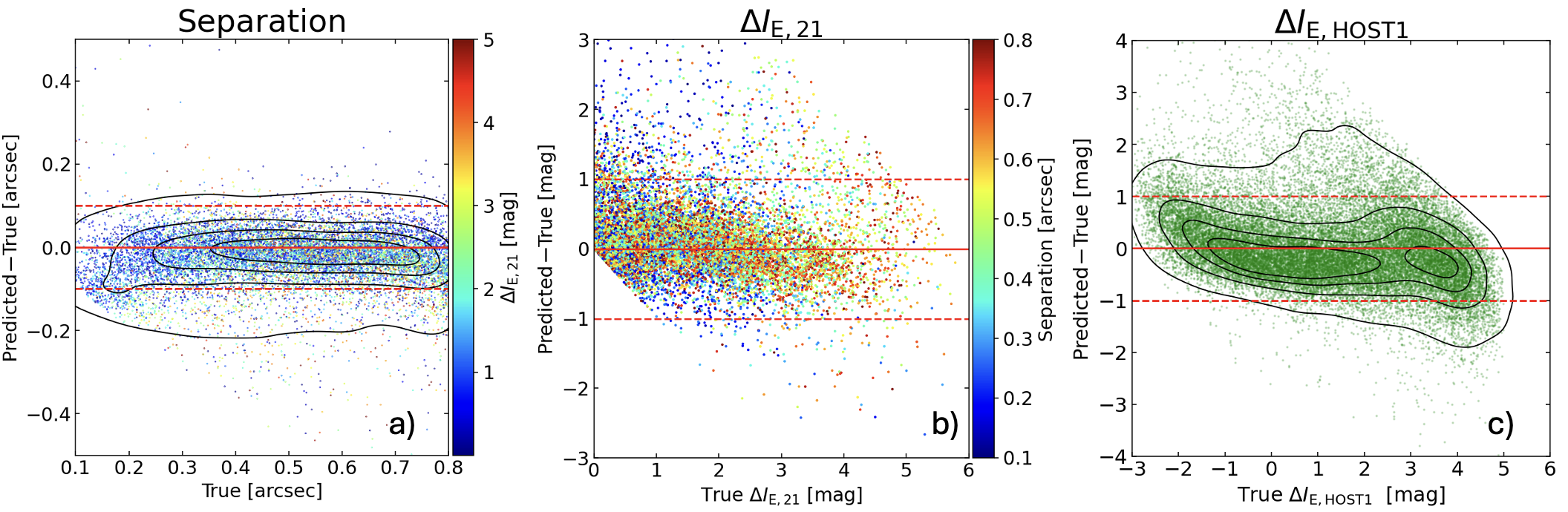}
    \caption{Prediction of the CNN of objects classified as dual AGN. Scatter plots comparing true values (\textit{x}-axis) with the difference of the predicted values and the true values (\textit{y}-axis) for different parameters derived from the CNN. (a) Predicted separation (in arcsec) between the two sources. The colour bar represents the $\Delta I_{\mathrm{E,21}}$ for the data points. (b) Prediction of the $\Delta I_{\mathrm{E}}$ between the secondary and primary sources ($\Delta I_{\mathrm{E,21}}$). The colour bar represents the separation in arcsec of the two sources. (c) Prediction of the magnitude difference ($\Delta I_{\mathrm{E}}$) between the host and primary source $\Delta I_{\mathrm{E,H1}}$. Red dashed lines indicate perfect predictions and $\pm$0.1 in arcsec error margins ($\pm$1 in px) and  $\pm$0.5 in mag.  }
    \label{fig:prediction} 
\end{figure*}

We evaluated the performance of the CNN by analysing its receiver operating characteristic (ROC; \cite{Bradley1997}) curve. This curve is constructed by varying the threshold probability used to classify galaxies as dual or single, considering the number of True Positives (TPs), True Negatives (TNs), False Positives (FPs), and False Negatives (FNs). The TP rate, also called completeness, which measures the fraction of true mergers correctly identified, and the FP rate, which indicates the fraction of single AGN incorrectly classified as dual, are computed as follows:

\begin{align}
    \text{TPR} &= \frac{\text{TP}}{\text{TP} + \text{FN}},\label{tpr} \\
    \text{FPR} &= \frac{\text{FP}}{\text{FP} + \text{TN}}.
    \label{fpr}
\end{align}

Figure \ref{fig:roc_con_host} shows the ROC curves for the test set. 
We used a one-vs-rest (OvR) approach, defining classes 1 and 2 as the positive class and class 0 (single sources) as the negative class. The probability of the positive class was then computed as the sum of the predicted probabilities for classes 1 and 2.
The perfect classifier reaches the top left corner, with the maximum TP rate and a FP rate of 0. Hence, we can use the area under the curve (AUC) as a metric of performance. The AUC curve in our CNN reaches 0.98.
We want to compare the performance of our CNN with the performance of standard methods such as \texttt{SExtractor}\footnote{\url{https://sextractor.readthedocs.io/en/latest/Introduction.html}} or the \texttt{IRAFStarFinder} module of \texttt{photutils} \citep{Bradley2021}.
The deblending method adopted in SExtractor is based on multi-thresholding and works on any kind of object but it is unable to deblend components that are so close that no saddle is present in their profile.
A better optimiser for point source detection is \texttt{IRAFStarFinder} that searches images for local density maxima that have a peak amplitude greater than a threshold above the local background and have a PSF FWHM similar to the input FWHM. 
Therefore, we measured the ROC curve of the standard method with \texttt{IRAFStarFinder} increasing  5, 10, 15, 20, 30~$\sigma$, and with a minimum detection separation from the limit imposed by the algorithm (2$\times$FWHM). Obviously, these values are simply for the purpose of having a direct comparison between the two curves. 
We first calculated the true positives by applying the algorithm to all simulations with dual objects and then the false negatives by applying the algorithm to the single objects. For the calculation of TP, we considered the correct detection if the algorithm finds at least one source within \ang{;;0.8} of the primary source. Detection of sources at larger distances this means that the network has detected other source present in the field (foreground objects). We note that this represents an upper limit to TPs since we consider a correct detection even if the algorithm detects possible contamination within \ang{;;0.8} (clumps, structure in the galaxy). For the calculation of FNs, we applied the algorithm to single sources considering an incorrect prediction if it finds more than one source within \ang{;;0.8}.
The ROC curves of the neural network and \texttt{IRAFStarFinder} predictions are shown in Fig.~\ref{fig:roc_con_host} as red line and with blue dots for increasing sigma, respectively. 
The ROC curve of the CNN lies above that of \texttt{IRAFStarFinder}, indicating that the CNN achieves higher completeness and precision across all detection thresholds. Classical methods do not reach the same level of completeness as the CNN without substantially increasing the number of false positives.


\subsection{Confusion matrix}
Figure \ref{fig:conf_matrix} shows the confusion matrix, where the \textit{y}-axis represents the true labels and the \textit{x}-axis represents the predicted labels. 
The labels are defined as follows: 0 corresponds to a single AGN, 1 to a dual AGN with separation < \ang{;;0.25}, and 2 to a dual AGN with separation larger than \ang{;;0.25}.
If we consider dual candidates labelled as 1 or 2, we find 337 false positives, corresponding to approximately 2\%, and a completeness of about 78\%. This leads to a F1 score, defined as the harmonic mean of the precision and completeness (recall) scores, of 0.86. If, instead, we restrict the analysis to candidates labelled as 2 (dual AGN), the precision increases to 98.6\%, with a completeness of 68\% and a F1 score of 0.80. However, these global metrics are not necessarily meaningful in a real case scenario since the true distributions of source separations and magnitudes are unknown. In particular, in the limiting case where all dual AGN lie at separations $\sim \ang{;;0.25}$, where some models predict the peak of their number density \citep{Saeedzadeh2024}, the completeness metric would lose significance, since many of them would not be detected.
What is more important is to understand how precision and completeness vary across the physical parameter space.

\begin{figure}
    \centering
    \includegraphics[width=1\linewidth]{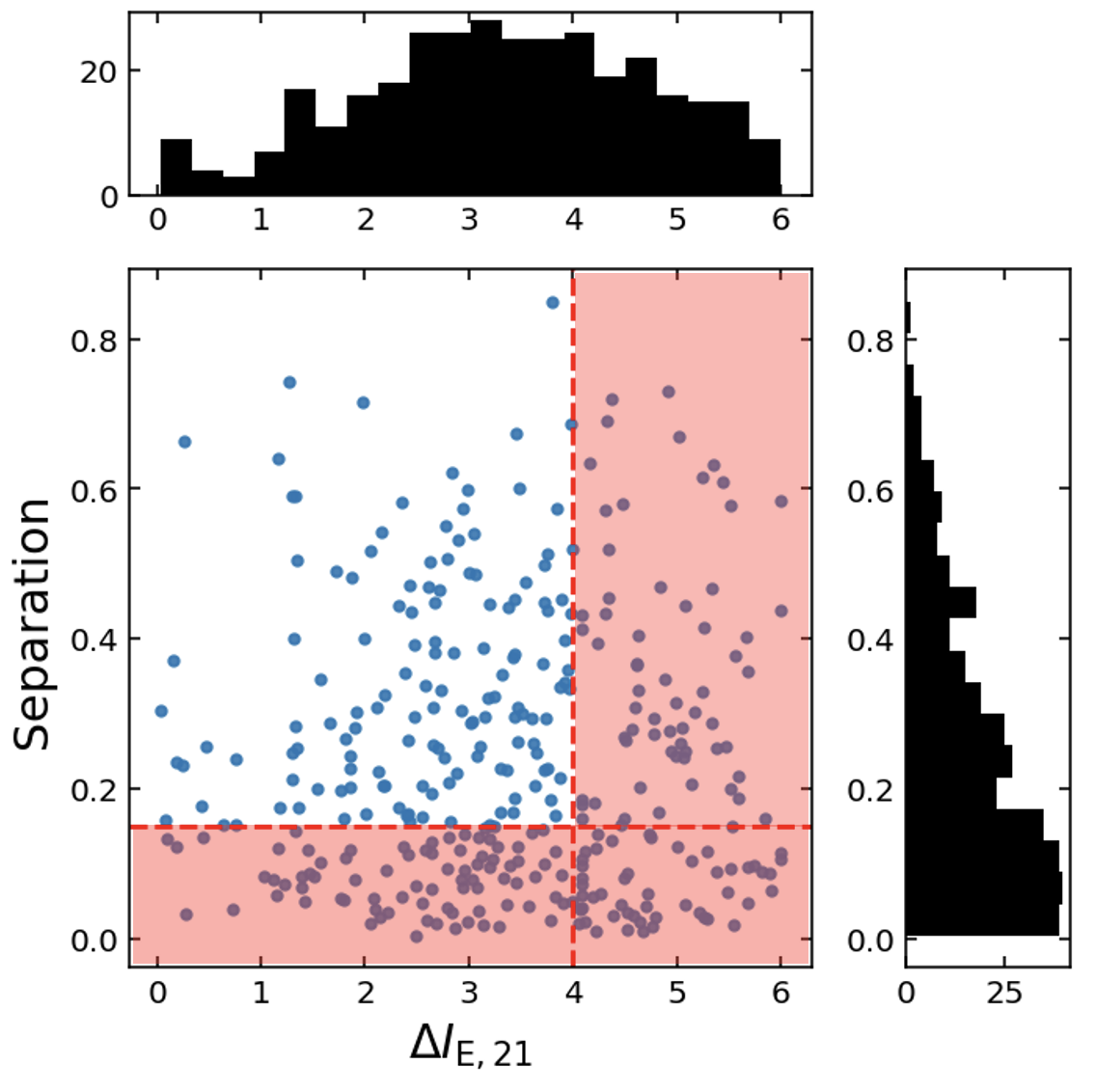}
    \caption{CNN prediction of separation and  the magnitude difference between the primary and secondary sources ($\Delta I_{\mathrm{E,21}}$) for FP objects. The top and right histograms show the distribution of $\Delta I_{\mathrm{E,21}}$ and of the separation of FPs, respectively.  The red shaded regions indicate  the excluded objects, defined by a separation smaller than \ang{;;15} and $\Delta I_{\mathrm{E,21}}$ higher than 4.   }
    \label{fig:FP Threshold}
\end{figure}

\begin{figure}[t!]
\centering
    \includegraphics[width = 0.46 \textwidth]{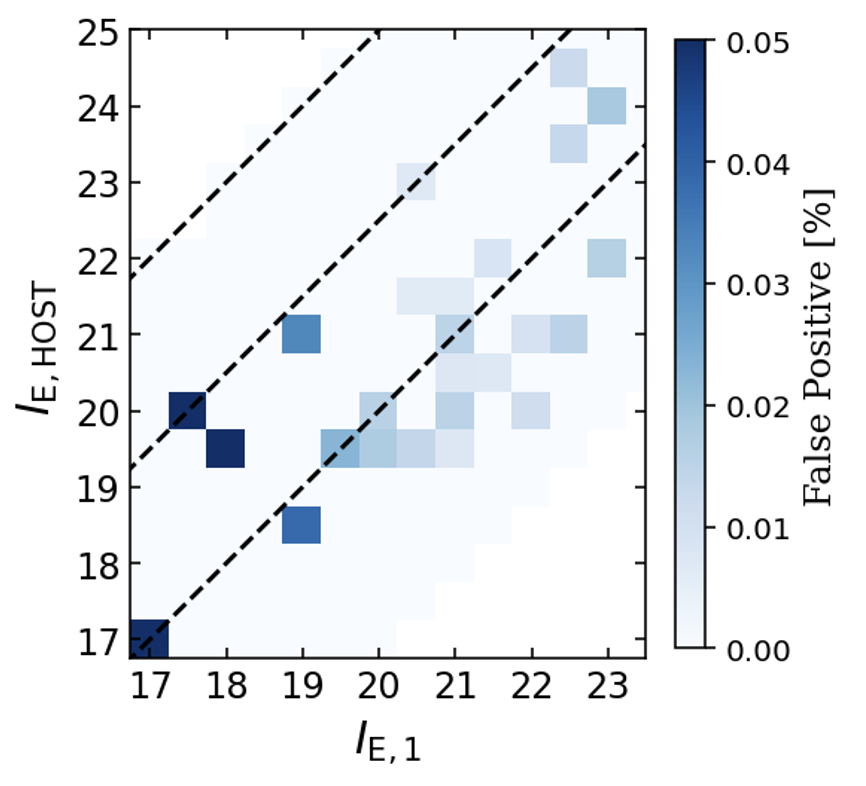}
    \caption{FP rate in bins of magnitude of the host galaxy, $I_{\mathrm{E,HOST}}$, and of magnitude of the primary AGN, $I_{\mathrm{E,1}}$.}
    \label{fig:precision}
\end{figure}

\begin{figure*}[t!]
\centering
    \includegraphics[width = 1 \textwidth]{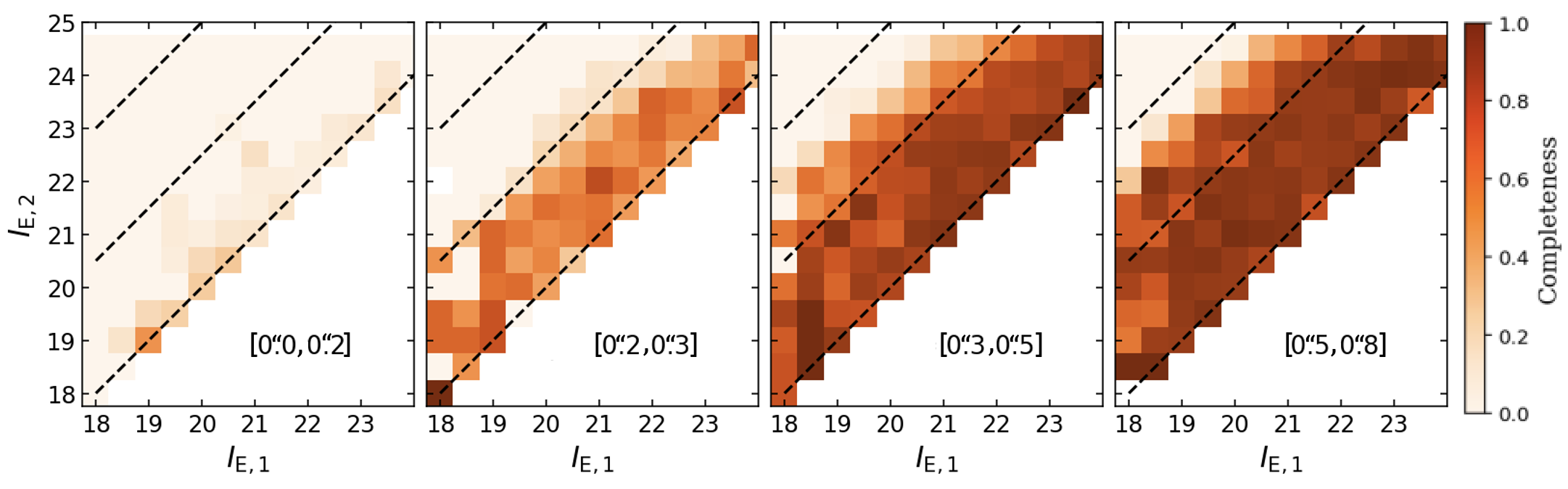}
    \caption{Completeness in bins of $I_{\mathrm{E,1}}$ and $I_{\mathrm{E,2}}$ for different separation ranges. The black dashed lines represent flux ratios of 1, 10, and 100 between the primary and secondary sources. }
    \label{fig:accuracy_detection}
\end{figure*}

\subsection{Predictions of the model}
 As shown in Fig.~\ref{fig:CNN}, the network has four different end layers that predict the nature of the AGN, but also the exact location of the secondary source, $\Delta I_{\mathrm{E,21}}$, $\Delta I_{\mathrm{E,H1}}$. 
Figure \ref{fig:prediction} shows the network predictions, where the x- and y-axis are respectively the true value, and the difference between the network prediction and the true value.
Panel a) presents the predictions of the separation between the two objects computed as $d = \sqrt{(x_0 - x_{\mathrm{pred}})^2 + (y_0 - y_{\mathrm{pred}})^2}$ where $x_{0}$ and $y_{0}$ are the central pixel of the primary source, while $x_{\mathrm{pred}}$ and $y_{\mathrm{pred}}$ are the coordinates of the secondary source predicted by the CNN. The plot only includes sources that are predicted to be dual sources. When the network is not able to detect a secondary component, it also cannot accurately assign a distance, increasing the scatter in the plot.
We see that the prediction is correct within $\pm\ang{;;0.1}$. 
Panels b) and c) show  $\Delta$\IE between the secondary and primary source ($\Delta I_{\mathrm{E,21}}$) and $\Delta$\IE between the host and primary ($\Delta I_{\mathrm{E,H1}}$). The prediction of $\Delta I_{\mathrm{E,21}}$ shows the scatter from the true value of $\sim~\pm$0.5 mag that increases when separations are small. This is mainly due to the fact that, at close separations, the network can no longer perform precise photometry. 

\subsection{Threshold to filter}\label{sec:threshold}
In order to choose the best thresholds for filtering, it is also necessary to consider the number of dual AGN that we expect. According to simulations the number of dual AGN can vary between 0.1 and 10\% of all AGN \citep{Steinborn2016,Rosas-Guevara2019,Volonteri2022,PuertoSanchez2025}, so being a class of rare objects, it is necessary to choose a metric such that the FPR are minimised compared to TPR.
To do this, the output of the neural network can be used to filter out objects that most likely represent contaminants. 
We try to understand how completeness and precision vary as a function of network output by changing the thresholds of the output. 
In particular, we vary the threshold of each parameter for the prediction of all dual systems in bins of magnitude of the primary, secondary, and separation. This allows us to estimate the expected completeness in real observations for each region of parameter space. We perform the same analysis for systems labelled as single, in bins of magnitude of the primary and of the host, to estimate the number of false positives.
For each trained network and for each value of the decision threshold applied to the network output, we calculate the TPR and FPR with Eqs.~(\ref{tpr}) and (\ref{fpr}).



To determine the optimal classification threshold, we adopted a modified selection metric defined as

\begin{equation}
\left( \text{TPR} \; f_{\text{dual}} - \text{FPR} \; f_{\text{single}} \right) \; N_{\text{QSO}},
\end{equation}
where \( f_{\text{dual}} \) and \( f_{\text{single}} = 1 - f_{\text{dual}} \) represent the assumed fractions of dual and single AGN, respectively, and \( N_{\text{QSO}} \) is the total number of QSOs in the input sample. This metric approximates the expected number of true positives minus false positives, and was evaluated in the separation range of \ang{;;0.2} to \ang{;;0.6}, where the number of dual AGN is expected to be higher. Assuming a dual AGN fraction of 1\% and \( N_{\text{QSO}} = 6000 \), the resulting optimal thresholds is 0.9 and typically yield an expected net number of true detections (TP $-$ FP) between 10 and 20. However, this estimate relies on several assumptions, in particular that the separation and magnitude distributions of dual AGN in the real data mirror those used in our simulations, an approximation that may not reflect reality.

\subsubsection{Investigation of false positives}\label{sec:FP}
We investigated the behaviour of the CNN in relation to false positives, that are sources labelled as single but predicted to be class 1 or 2 by the network. Figure~\ref{fig:FP Threshold} shows the CNN predictions for $\Delta I_{\mathrm{E,21}}$ and separation for the 337 false positives identified out of a total of 15\,969 sources. 
Some of these FPs represent foreground contamination that may be present because we used stamps of real galaxies in which there is a non-zero probability that a foreground source is present. Another kind of FPs are galaxies with disturbed morphology, that may be misclassified.
To better isolate probable FPs, we applied a posteriori physical threshold, selecting only those with $\Delta I_{\mathrm{E,21}} < 4$
and separation $<\,$\ang{;;0.15}. This additional cut allows us to eliminate more than half of the FPs though at the cost of also removing some TP. This choice is justified by physical considerations and helps to improve the overall robustness of the classification.

We studied the classification accuracy as a function of only the parameters available for single AGN: the magnitude of the primary source and that of the host galaxy. As shown in Fig.~\ref{fig:precision}, the overall precision decreases when the host galaxy is particularly bright although it remains high (i.e., produces a low FP rate) with the chosen threshold. 


\subsection{Detection accuracy}

We investigate how the completeness varies as a function of physical parameters, specifically the magnitudes of the primary and secondary components and their separation, using the selected thresholds. We apply the CNN with a classification threshold of 0.9, a minimum separation of $0\farcs15$, and  $\Delta I_{\mathrm{E,21}} < 4$ as described in Sect.~\ref{sec:FP}.

Figure~\ref{fig:accuracy_detection} shows the resulting completeness across this parameter space. At separations between $0\farcs2$ and $0\farcs3$, the CNN achieves an average completeness of approximately 60\% in classifying dual sources with a contrast of up to a factor of ten. For larger separations, dual objects with $\Delta I_{\mathrm{E,21}} < 4$ are detected with very high completeness.
However, at separations greater than $0\farcs5$--$0\farcs6$, standard deblending algorithms, such as \texttt{IRAFStarFinder}, perform better in terms of detection accuracy, but as described in Sect.\ref{sec:ROC}, they do not reach the performances of the CNN.

\section{Application to Q1 data}\label{sec:applicationtoq1data}
Starting from the initial dataset of approximately 15\,000 AGN cutouts, we applied a series of selection criteria to refine the sample. We restricted the \IE range of the primary source to 18–23.5 and ISOAREA < 450, where the ISOAREA is defined as the area (in number of pixels) above the detection threshold, in order to remove very extended sources, and removed duplicate entries. Moreover, we excluded any QSO whose coordinates did not correspond to a detected source within \ang{;;0.3}. After applying these cuts, we ended up with 5993 cutouts, a sample size that is sufficiently  to robustly test our method and to provide a realistic estimate of the number of dual systems to be followed-up with spectroscopic observations.


\subsection{Application of the CNN}
We first preprocessed the images by normalising the pixel values between 0 and 1. Each image was then centred on the pixel with the maximum flux, following the same procedure used during the training of the CNN on the simulated data (see Sect.~\ref{Sample}). Finally, we applied the neural network on the preprocessed images.
We found 265 candidates classified as dual systems. Using the threshold defined in Sects. ~\ref{sec:threshold} and Sect.~\ref{sec:FP}, we identified a sample of 49 objects that meet the selection criteria, representing the $\sim\,$0.8\% of the initial sample. In this selected candidates, given the high precision of the neural network, we aimed to minimise contamination from objects with bright host galaxy structures. 
The main contaminants are represented by foreground sources, such as stars, and it is therefore necessary to estimate the expected number of such contaminants. The distribution of the separations predicted by the neural network is shown in Fig.~\ref{fig:contamination}.

\subsection{Foreground contamination}\label{sec:foreground_cont}
Quantifying contamination from foreground stars within our candidate pair sample is inherently challenging. In this section, we provide a statistical estimate of the expected number of chance alignments with foreground sources along the line of sight, compared to systems that are more likely to be physically associated with the primary source, such as dual or lensed AGN or clumpy structure in the galaxy.

To estimate the probability of contamination from unrelated sources, we performed a simple statistical test by making random square cutouts across the analysed fields of view. Specifically, we extracted 20\,000 random $2'' \times 2''$ cutouts and ran \texttt{SExtractor} on each of them. The size of the cutouts was chosen to avoid truncating sources at separation of around \ang{;;1}. The estimated number of contaminants depends on both the limiting magnitude and the separation from the center within which sources are classified as contaminants. For example, by restricting to \IE $<$ 25 and separations $<$ \ang{;;0.8}, \texttt{SExtractor} finds 605 contaminants. When extending the limits to \IE $<$ 26 and separations $<$ 1\arcsec, the number increases to 1650 out of 20\,000, corresponding to approximately 8\% contamination.
If we consider only those with CLASS STAR > 0.5, which would represent the true contaminants of dual or lensed AGN, \texttt{SExtractor} identifies 82 (0.4\%: 24 out of $\sim\,$6000) and 175 (0.87\%: 52 out of $\sim\,$6000) contaminants in the two respective thresholds.

We also provide estimates of the contaminants using the official catalogue provided by the \Euclid pipeline (MER catalogue). For each MER tile catalogue, we randomly generated a set of RA and Dec positions uniformly distributed within the boundaries defined by the minimum and maximum RA and Dec of the tile. For each generated coordinate, we associated the closest object in the catalogue. We repeated this procedure 100\,000 times. Under the previous thresholds, we find $\sim\,$7\% contamination for \IE < 26 and separation < 1\arcsec, and $\sim\,$3\% for \IE < 25 and separation < \ang{;;0.8}. 
When using PROB POINT LIKE (the estimation uses a “star probability cube”) in the MER we found only 169 sources out of 100\,000 images (0.17\%). 





It is important to note that all of the contaminant sources identified in our statistical test are detectable by the source extraction algorithm, as this method does not depend on the presence of a primary object. However, the most important question for our analysis is: how many of these contaminants would actually be detected by our CNN? 
Another noteworthy aspect is that, while it is possible to make sharp cuts in magnitude and separations for the contaminants when using \texttt{SExtractor}, the dual candidates (in this case, foreground contaminants) found by the CNN do not necessarily meet the criteria for magnitude and separation. This is because the CNN was trained with normalised images between 0 and 1 and with secondary sources at separations up to \ang{;;0.8}. Thus, it is not possible to include a cut in magnitude of the secondary, and most likely, it is not certain that the network only detects sources up to \ang{;;0.8}. Moreover, the network has been trained to detect point-like secondary objects (secondary AGN or lensed QSO), so it is not clear a priori how the network behaves with extended contaminants (e.g., galaxies). To estimate how many contaminants we expect from the neural network, we added to the random cutouts a point source with a \IE between 18 and the \IE of the contaminant (if present) following the same distribution of our input QSO sample.  
We compare the number of contaminants estimated by \texttt{SExtractor} with the number of contaminants estimated by the CNN. The network finds in total 118 contaminants out of 20\,000 images, so around 30 contaminants out of the initial sample of $\sim\,$6000 QSOs. We note from Fig.~\ref{fig:foreground_cont} in Appendix \ref{appendix:foreground} that, as expected, the network tends to consider as contaminants mostly systems with point source contaminants. 

\begin{figure}
    \centering
    \includegraphics[width=1\linewidth]{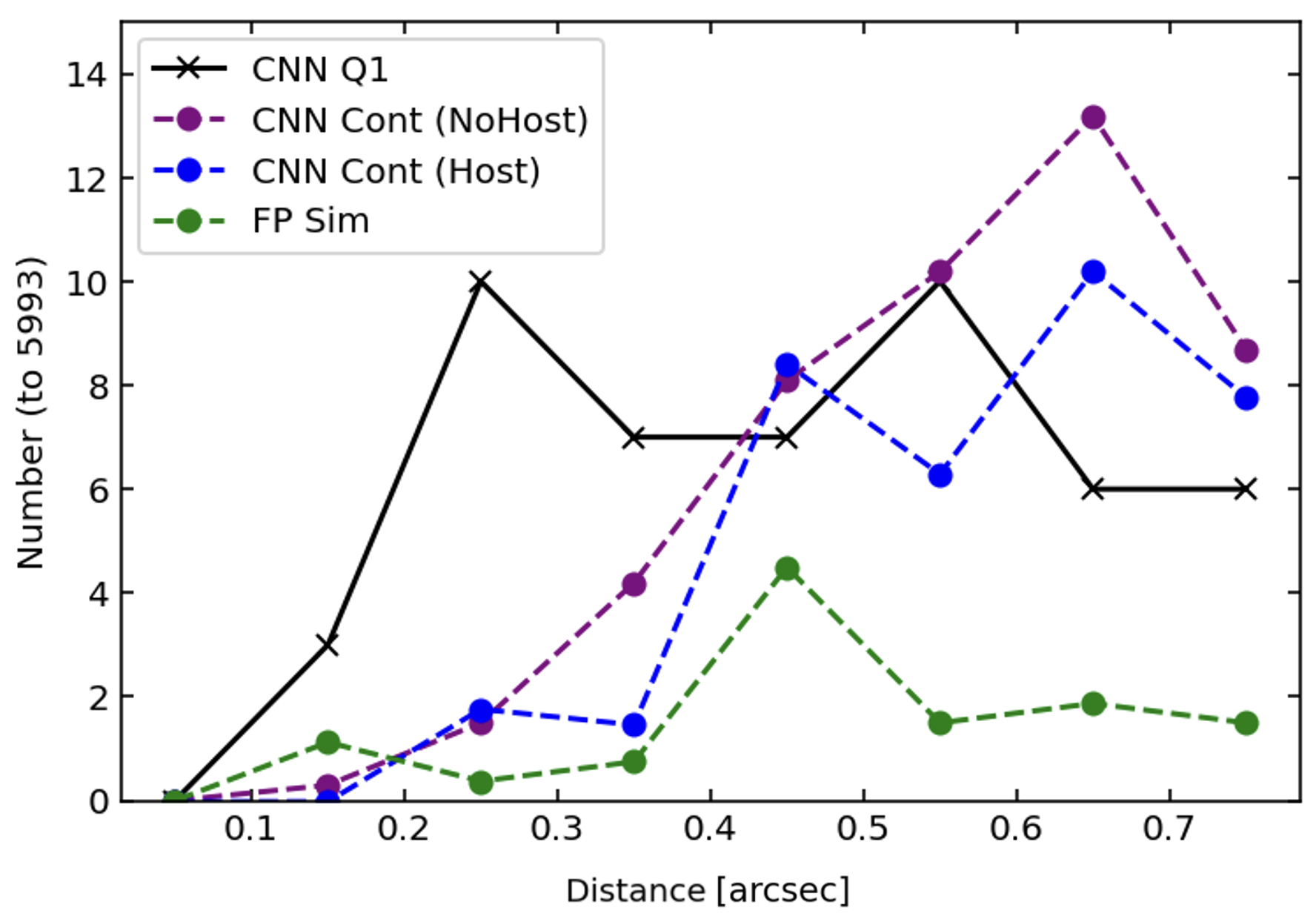}
    \caption{Comparison between the number of dual candidates found by CNN and the expected contaminants. The number is normalized by the total number of AGN considered after the cuts (5993). Dual AGN candidates found by CNN in bins of separation (in black), number of expected contaminants classified dual by the CNN once we added a primary source without (purple) and with (blue) host galaxy, and
    false positives expected from simulations (in green).}
    \label{fig:contamination}
\end{figure}

In the real case, the presence of a galaxy host can significantly affect the detectability of a nearby secondary source, especially when the host is brighter than the companion. In particular, faint secondary sources are more easily detected when no host is present. As a result, the number of contaminants estimated from random cutouts tends to exceed the number of pair sources identified around primary QSOs in the Q1. This difference is due to the presence of host galaxy structure in real systems, which lowers the efficiency of detecting nearby companions. We then included a host galaxy in the simulated cutouts, modelled with a Sérsic profile and a magnitude ranging between  $ I_{\mathrm{E,1}}$=–1 to $ I_{\mathrm{E,1}}$=+4 and convolved with the PSF. When applying the CNN to these images, we find that the number of contaminants is reduced compared to the case without a host galaxy (blue line in Fig.~\ref{fig:contamination}).



\begin{figure*}[t!]
    \centering
    \includegraphics[width=1\linewidth]{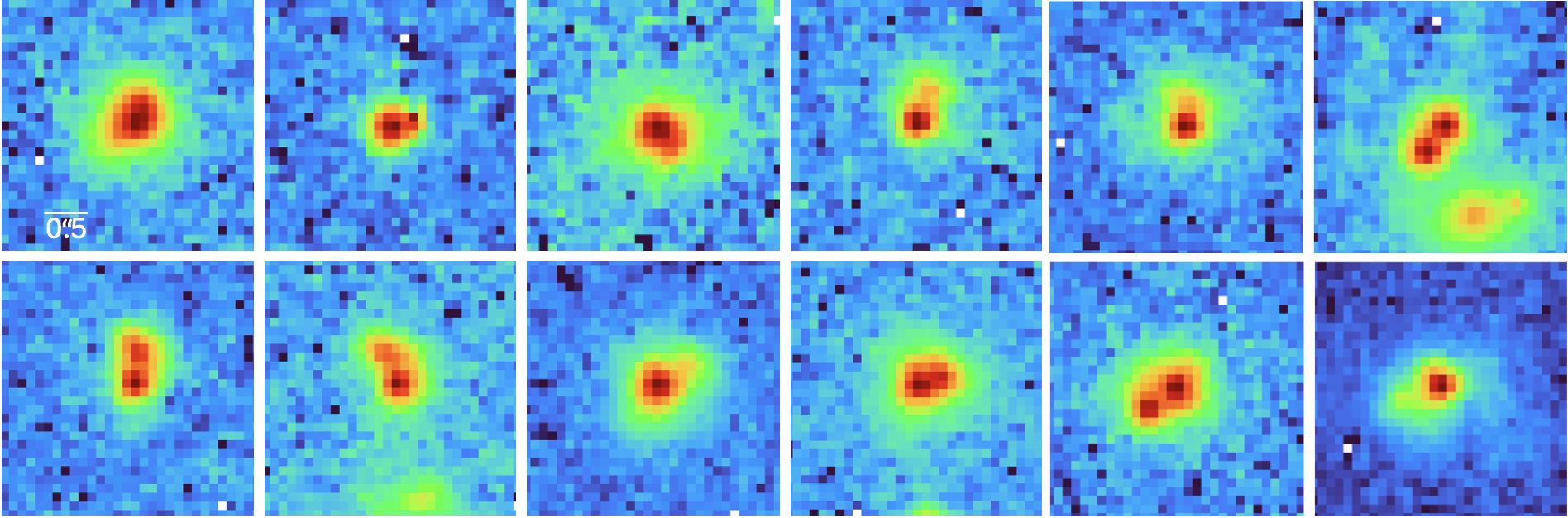}
    \caption{Dual AGN candidates found by the network with separations between \ang{;;0.15} and \ang{;;0.45}. The number of objects in this separation range exceeds the number expected from foreground objects and false positives caused by clumpy galaxy structure, making these candidates more reliable than those at larger separations. The size of each cutout is $\ang{;;3} \times \ang{;;3}$, and the flux is shown in logarithmic scale.   }
    \label{fig:Closesources}
\end{figure*}

Another important aspect involves the number of expected FPs from simulations. While the network has shown high precision overall, we observed that in some cases, single systems in simulations were misclassified as dual systems. To account for these different contaminants, we include the histogram of predicted distances for FPs in the simulations, shown in green in Fig.~\ref{fig:contamination}. In other words the green line represents the FPs that we expect from simulations.
We note that the dual sources identified in the Q1 sample by the network with separations of \ang{;;0.2}–\ang{;;0.4} exceed the number expected from foreground objects and false positives caused by clumpy galaxy structure. This excess is likely indicative of a population of dual candidates (see Fig.~\ref{fig:Closesources}). At larger separation (>\,\ang{;;0.5}), the number of contaminants is consistent with the number of detected double sources. While this does not rule out the presence of dual AGN candidates at these higher separations, the probability of finding good candidates with secondary sources brightest than 25 mag is lower. As we address in Sect. \ref{sec:PSF_sub}, at larger separations it is possible to exploit the four \Euclid bands to give a probability that the secondary source is an AGN.


\begin{figure}
    \centering
    \includegraphics[width=1\linewidth]{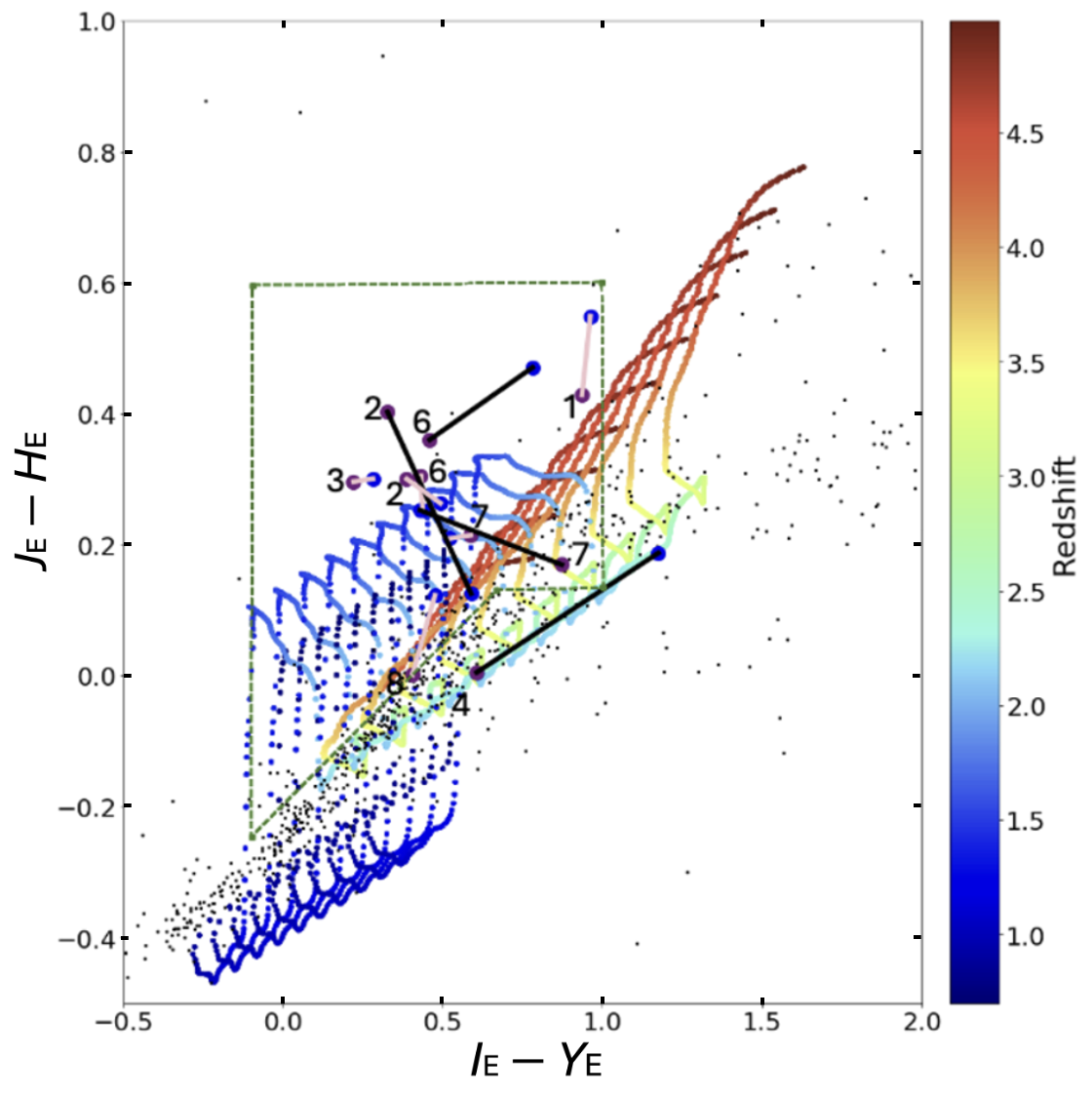}
    \caption{Colour-colour diagram of dual AGN candidates in Fig.~\ref{fig:phot_cand} overlayed with simulated AGN colour coded with redshift \citep{Temple2021} and stars (black dots). The numbers correspond to the sources shown in Fig.~\ref{fig:phot_cand}. Each line connects the primary (purple dots) to the secondary (blue dot) source in a dual AGN candidate. Pink lines indicate photometry derived from MER catalogue using a 1~FWHM aperture, while black lines correspond to photometry obtained after PSF subtraction. The green dashed box indicates the AGN selection region defined in \cite{Q1-SP027}.}
    \label{fig:phot_colour_colour}
\end{figure}

\begin{figure*}[t]
    \centering
    \includegraphics[width=1\linewidth]{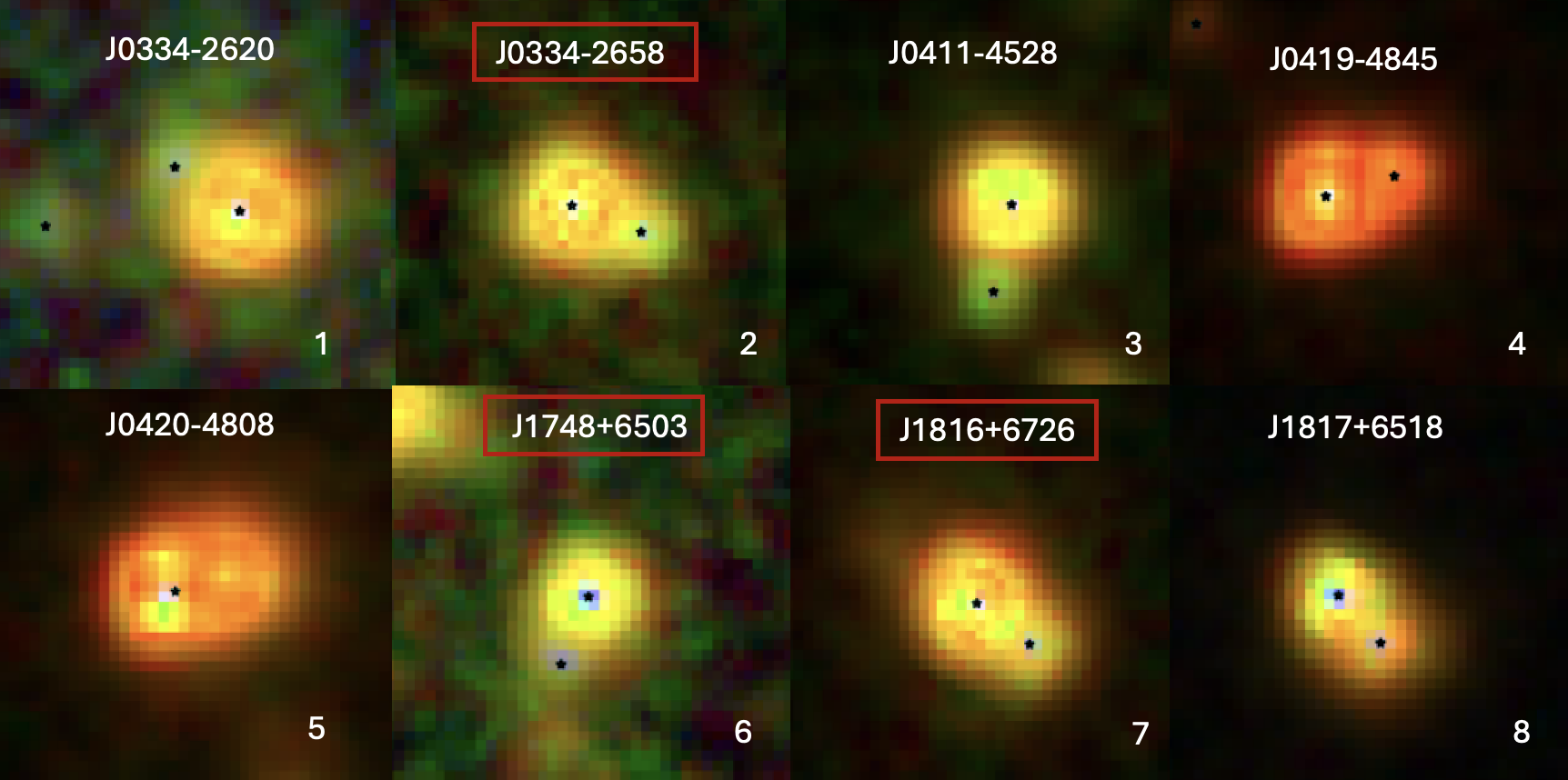}
    \caption{Three-colour composite images (\IE, \JE, \HE) of dual AGN candidates derived from MER photometry or PSF subtraction. Systems enclosed in red boxes are those where both sources lie within the AGN region in both photometric methods. The black stars mark the positions of the input sources from the MER catalogue. The size of each cutout is $\ang{;;3} \times \ang{;;3}$.}
    \label{fig:phot_cand}
\end{figure*}


\subsubsection{Photometry and PSF subtraction}\label{sec:PSF_sub}
\Euclid obtains images in four bands as described in Sect.~\ref{Eucliddata}, and the colours can in principle be used to study the nature of the sources.
For systems with multiple entries in the MER catalogue, we can use MER fluxes to study the photometry with different apertures. 
To ensure reliable colour measurements, we implemented our own photometric pipeline, as accurate PSF subtraction is essential when analysing closely separated sources. While the MER catalogue provides photometric measurements using a range of aperture sizes, these are often affected by flux contamination of close sources, which is the case for our candidates. 
To perform the photometry on the candidate sample, we implemented a customized PSF-fitting algorithm for measuring the photometry of the blended sources. We first used \texttt{FittableImageModel} of \texttt{photutils.psf} on the VIS PSF of the MER catalogue mosaic at the source position. Then we used \texttt{IRAFStarFinder} on the VIS stamp, which has higher resolution, to search for point sources. Once we found the positions of the two sources, we measured the fluxes in the VIS and NISP bands by using the appropriate PSF (from MER) and fixing the positions to within 1 pixel. We also included in the fit a S\'ersic component convolved with the PSF of the corresponding band to model the host galaxy. The effective radius ($r_\mathrm{eff}$) was allowed to vary between 2 and 3 pixels, the S\'ersic index ($n$) between 2 and 4,  the ellipticity in the range 0 and 1, and the position within a radius of 1 pixel from the center.
We then summed the fluxes of each pixel in the model and using the ZERO MAG, we computed the magnitude of the two sources and of the host galaxy in each band.

\subsubsection{Colours of the companion candidates}
Figure~\ref{fig:phot_colour_colour} shows a colour-colour diagram (\(\JE - \HE\) vs. \(\IE - \YE\)) used to analyze the photometric properties of dual AGN candidates. The coloured points represent simulated AGN generated using the \texttt{qsogen} package from \cite{Temple2021}, with redshift varying between 0.7 and 5 and extinction $E(B-V)$ ranging from 0 to 0.1. Black points correspond to stellar templates from the \texttt{XSL\_DR3\_release} catalogue \citep{Verro2022}, included to trace the stellar region in colour space. The green dashed box indicates the AGN selection region defined in \cite{Q1-SP027}, which is used to separate AGN from stars. For each dual AGN candidate pair, we compute and plot the colours of both the primary and secondary components. 
We measured the colours of all source pairs for which reliable photometry could be performed for all bands. In general, for dual sources with separations $>$ \ang{;;0.5}, we obtain good fits with errors on the flux of the two sources about 20\%, which corresponds to an error on the magnitude of $\sim\,$0.2 mag. To assess the consistency of our photometric method, we compared the colours derived from our PSF subtraction technique with those listed in the MER catalogue, where the two sources are recorded as distinct entries. Figure~\ref{fig:phot_cand} presents seven source pairs in which both components fall within the AGN colour region in at least one of the two photometric methods (PSF subtraction or MER catalogue) and one source (J$0420-4808$) where the colours of both sources are consistent with being at the same redshift ($z\sim$~2.5). Such a comparison is only feasible for pairs with angular separations larger than \ang{;;0.5}. Only three dual systems exhibit both components within the AGN region (red box in Fig.~\ref{fig:phot_cand}), consistent with previous findings suggesting that most widely separated pairs are likely due to line of sight (LOS) contaminations rather than physically associated AGN.

\subsubsection{Visual inspection}
\label{sec:visual-inspection}
We conducted a visual inspection of the QSO sample input to the network, categorising objects as follow:

\begin{itemize}
    \item[0.] Clear dual objects: These are well-separated pairs where both components are distinctly visible and compatible with being point sources. 
    \item[1.] Possible duals at low separations: Objects that could be dual sources but are at low separation (<\,\ang{;;0.25}) or with asymmetric morphology, making identification less certain.
    \item[2.] Contaminated objects: These include secondary sources that are not point-like sources or that are point-like but have a separation greater than \ang{;;0.8}.
    \item[3.] Possible lens-like structures: Sources that exhibit morphological features like gravitational lensing effects such as arcs.
    \item[4.] Single objects: Sources with no secondary components visible.
    \item[5.] Objects with a bright host: QSOs where the presence of a luminous host galaxy that could affect the classification.
\end{itemize}

Out of the 5993 target in the input sample, we found 51 clear dual objects (class 0), 40 possible dual objects at low separation (class 1), about 340 classified as contaminations, 8 possible lenses, 5208 singles, and 336 associated with bright hosts. We compared the dual AGN candidates found by the CNN with the visual inspection. Thirty-two candidates belong to class 0 or 1, 8 belong to class 2, two are possible lenses (see Fig.~\ref{fig:possible_lens} in Appendix~\ref{appendix:lenscandidates}), and 7 belong to class 4 or 5. We note that visual inspection was carried out in log scale on images normalized between 0 and 1, so we classified as dual objects with a very weak secondary when the network was trained to detect secondaries as faint as \IE = 25. Table~\ref{tab:threshold_metrics} shows the number of TPs and FPs based on visual inspection identified by the CNN across different threshold values. TPs and FPs refer to the class identified during the visual inspection (P: class 0, 1; N: other classes). The configuration adopted in this work is the first one which, as suggested by simulations, maximizes the difference between the TPs and FPs. For comparison, the other two configurations maximize  TP/FP and the number of TPs but significantly increasing the number of FPs.
We also applied visual inspection to simulations, finding that we were able to correctly classify only 4\% of simulated dual objects between ~\ang{;;0.15} and ~\ang{;;0.2}, while between ~\ang{;;0.25} and ~\ang{;;0.3} we classify correctly 60\% of objects.


\subsection{Additional space-based imaging}
\begin{figure}
    \centering
    \includegraphics[width=1\linewidth]{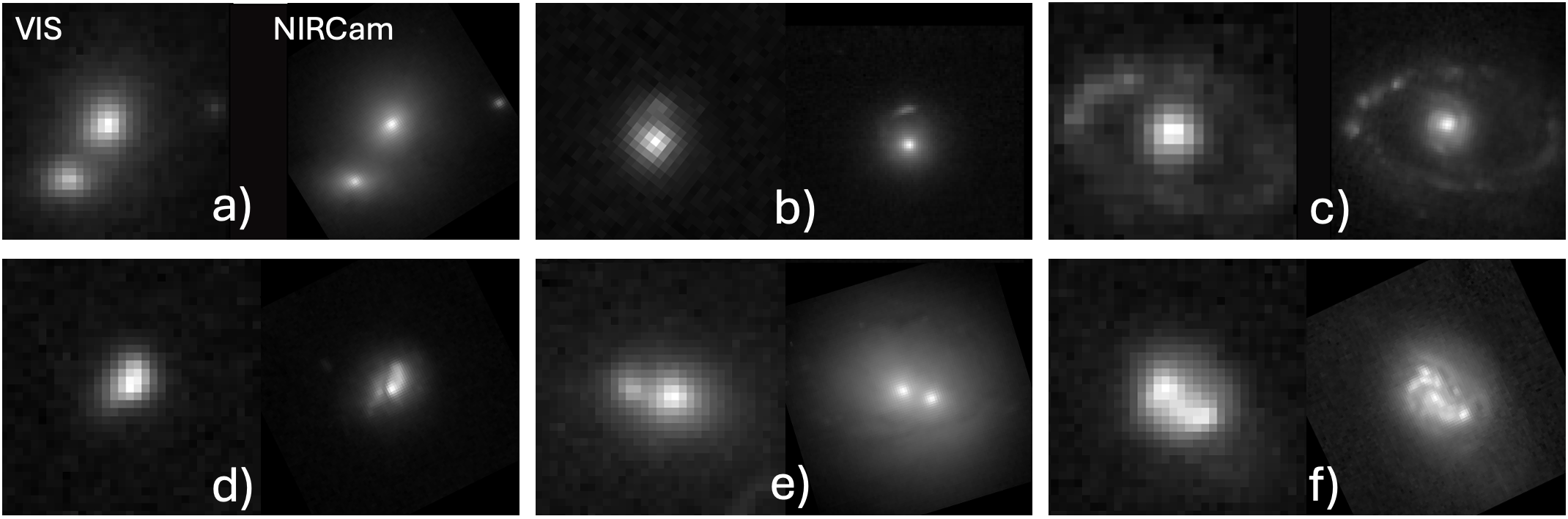}
    \caption{VIS and NIRCam/JWST images of objects predicted dual without considering thresholds.}
    \label{fig:high_res_facilities}
\end{figure}

To test our results, we searched for higher resolution images of our dual AGN candidates obtained  with facilities such as JWST and HST.
In this case, we applied the network without using any threshold in \texttt{ISOAREA} and we defined as dual candidates all the objects for which the sum of the predicted probabilities for labels 1 and 2 is higher than 0.5 (lower than the threshold we used in Sect. \ref{sec:threshold}). 
This approach allows us to identify cases that may represent true positives which were excluded by the CNN due to a too conservative threshold. Moreover, it provides a way to investigate potential FPs and understand the sources of misclassification and to assess whether the observed number of FPs is consistent with expectations from simulations. However, this comparison relies on the assumption that the simulated distributions of separations and the magnitudes of the primary, secondary, and host components accurately reflect the underlying real distributions, which may not always be the case.
We confirm that when the thresholds are not applied the network does not always select sources with point-like secondary components in the data, as expected. Two sources with high resolution images were excluded from the initial selection due to ISOAREA values higher than 450 (both having known redshifts below 0.6) though they show clear indications of dual morphology (a and e in Fig.~\ref{fig:high_res_facilities}). In one case, we identify an arc-like feature that could be both a foreground lensed galaxy or a tidal feature (b). In two other cases, the network misclassifies the sources, since it is confused by the clumpiness and the irregular structure of the host galaxy (c, f). Another object presents a more ambiguous scenario (d, shown either in Fig.~\ref{fig:Closesources}): it may be a dual AGN with a separation of $\sim\,$~\ang{;;0.2}, or alternatively a single AGN whose nuclear emission is covered by a dust lane, creating two distinct components. These examples illustrate some of the limitations of the network when dealing with complex morphologies and highlight the importance of high-resolution imaging for proper classification.



\begin{table*}[t]
\centering
\caption{Performance metrics for different threshold configurations which maximise TP-FP (first row), TP/FP (second row) and TP (third row). TP and FP refer to the class identified during the visual inspection (P: class 0, 1; N: other classes).}
\begin{tabularx}{\textwidth}{X X X X X X X X X X}
\hline\hline
Threshold & Sep & $\Delta I_{\mathrm{E,21}}$ & $\Delta I_{\mathrm{E,H1}}$ & TP & FP & TP $-$ FP & TP/FP & TPR (\%) & FPR (\%)  \\
\hline
$>\,$0.90 & $>\,$\ang{;;0.15} & $<\,$4 & $>\,$$-2$ & 32 & 7 & 25 & 4.5 & 35 & 0.12  \\
$>\,$0.80 & $>\,$\ang{;;0.20} & $<\,$4 & $>\,$1 & 20 & 3 & 17 & 6.7 & 22 & 0.05  \\
$>\,$0.5  & $>\,$\ang{;;0.0}  & $<\,$5 & $>\,$$-2$ & 48 & 141 & $-93$ & 0.34 & 53 & 2.5 \\

\hline
\end{tabularx}
\label{tab:threshold_metrics}
\end{table*}

\subsection{MER catalogue}
Working directly on images and building a network made specifically for the dual AGN search is of paramount importance because in many cases the MER catalogue merges two objects with small separation into a single catalogue entry. 
It is interesting to understand which is the minimum distance between two point sources that allows the MER catalogue to distinguish the two objects and thus list them as two separate entries. To answer this, we calculated for each source in the catalogue the distance to the nearest source and studied in detail the pairs of sources with separations less than ~\ang{;;0.8}. In particular, we studied the minimum distance as a function of the \IE and Kron radius. We found that the MER catalogue is not effective for binary sources at distances less than \ang{;;0.8} and with flux ratios > 2. We checked which dual candidates predicted by the neural network are associated with two different entries in the MER catalogue. Of the 32 dual visual inspected sources of class 0 or 1, only one third have double entries in the MER catalogue.


\subsection{Summary of the CNN appliacation to real data}
We found 265 candidates classified as dual systems by the CNN. Using the threshold defined in the previous Section, we identified a sample of 49 objects that fulfil the selection criteria, representing 0.8\% of the initial sample.
Of these 49 objects, 32 were classified according to the visual inspection  as class 0 or 1, eight as class 2, two as class 3, and seven as class 4 or 5, as defined in Sect. \ref{sec:visual-inspection}. We showed that most systems with projected separations larger than \ang{;;0.5} are likely to be LOS contaminants. However, \Euclid-band photometry allows us to further investigate the nature of these systems. Among the objects with separations greater than \ang{;;0.5} and available MER photometry, seven systems have both the primary and the secondary components falling within the AGN selection region defined by \cite{Q1-SP027}. After performing PSF subtraction and adding a host galaxy component when required, we found that only three of these systems lie within the AGN region. In addition to these three, the most reliable candidates are those with smaller separations. These selected systems are reported in Fig.~\ref{fig:Closesources}. All the dual candidates we have found, once we excluded foreground contaminants, represent around the 0.25\% of the initial sample.

\begin{table}[ht]
\caption{Sample Selection}
\centering
\begin{tabular}{p{5.5cm} r}
\hline\hline
\textbf{Category} & \textbf{Number of objects} \\
\hline
QSO catalogue & 15317 \\
Selection and avoiding repetition & 5993 \\
D/L AGN (CNN w/o threshold) & 265\\
D/L AGN (CNN w/ threshold) & 49 \\
\texttt{"} \quad foreground contamination & $\sim\,$35 \\
\texttt{"} \quad D/L AGN cand. > \ang{;;0.45} (phot) & 3 \\
\texttt{"} \quad D/L AGN cand. < \ang{;;0.45} & $\sim\,$12 \\
\texttt{"} \quad lens cand.  & 2 \\
\hline
\end{tabular}
\label{tab:sample_selection}
\end{table}

\section{Discussion}\label{sec:discussion}
Estimating the true fraction of dual AGN systems remains highly challenging, even when follow-up observations are available. Providing a robust estimate of the dual fraction is beyond the scope of this work. We note that the definition of dual fraction is not always the same in all works, and different definitions may lead to underestimation or overestimation.
In this section, we explore how our results compare with cosmological simulations, particularly the \textsc{Horizon-AGN} simulation \citep{Volonteri2022}, and with previous studies.

\subsection{Comparison with simulations}

The dual fraction, i.e., the number of multiple AGN over the total number of AGN, is one of the main predictions of the models of SMBH formation and BH-galaxy coevolution. Our results are still too preliminary to test these predictions, but it is interesting to compare the number of candidates we select with the results of these predictions. 

The Horizon-AGN simulation \citep{Volonteri2022} predicts a dual fraction of 5\% when considering AGN with bolometric luminosities $L_{\mathrm{bol}} > 10^{44}~\mathrm{erg~s}^{-1}$ and projected separations less than $30$~kpc. From this we can obtain an order-of-magnitude estimate of the expected number of dual AGN in our sample by assuming that the dual fraction does not depend on redshift and bolometric luminosity. This assumption is very simplistic and known not to be true in most cases \citep{Comerford2015,DeRosa2019}, but can be used to obtain a rough estimate of the expected numbers.
Given our parent sample of $\sim\,$6000 AGN, this implies an expected number of $\sim\,$300 dual AGN systems with the same selections.
In this study, we focused on small projected separation ($<\,$\ang{;;0.8}). This threshold excludes about 80\% of the dual AGN population predicted by simulations, significantly reducing our expected number of detections. Moreover, our selection requires optical magnitude 
\IE < 25, which limits our sample to $L_{\mathrm{bol}}>\,10^{44}\,\mathrm{erg}~\mathrm{s^{-1}}$ at $z=1$ and $L_{\mathrm{bol}}>\,10^{45}\,\mathrm{erg}~\mathrm{s^{-1}}$ at $z=3$. This corresponds to only $\sim\,$1.8\% of the dual AGN population found in \cite{Volonteri2022}. As a result, we expect to identify $5-6$ dual AGN within our sample, at redshifts up to $z \sim 3$.
It is important to note, however, that the Horizon-AGN simulations do not resolve dual AGN at physical separations below $\sim\,$4~kpc, corresponding to \ang{;;0.5}. This limitation implies that our estimate is a lower bound, particularly at small separations where the majority of dual AGN are likely to reside. In fact, the most reliable candidates found in this work are at separations between \ang{;;0.2} and \ang{;;0.45}.

This range of separations is sampled by ROMULUS25 simulations \citep{Tremmel2017, Tremmel2019} presented in \citet[Fig.~7]{Saeedzadeh2024}, \citet[Astrid]{Chen2023}, and  \citet[Magneticum]{Steinborn2016}, which show that dual AGN dominate at small separations.
~\cite{Rosas-Guevara2019} showed that there is not a clear trend in the distribution at small distances ($<\,$15~kpc), while the dual AGN fraction increases around 20 and 25~kpc. 


\subsection{Comparison with other works}

Numerous studies have searched for dual AGNs using different selection techniques, including imaging and optical spectroscopy \citep{Hennawi2006,More2016,Eftekharzadeh2017,Mannucci2022,Scialpi2024,Jiang2025}, X-ray observations \citep{Koss2012, Gross2019,Guainazzi2021,Sandoval2023} and radio surveys \citep{Fu2015,Rubinur2018,Gross2019}. Each of these approaches, however, is subject to selection biases. A major observational bias arises from the fact that dual AGNs in the late stages of galaxy mergers are often heavily obscured by dust \citep{Ricci2017,Koss2018,Barrows2023}. Another important limitation is due to the sensitivity and the angular resolution of the instruments employed, which probe different dual AGN populations and different physical scales. As shown in Fig. 24 of \cite{Pfeifle2024}, dual AGN candidates reported in the literature are scarce at both high redshift ($z >$ 0.5) and small separations ($<\,$5 kpc). These unexplored regions of parameter space can be effectively probed thanks to the capabilities of \Euclid. The dual AGN fractions reported in previous works are not directly comparable to our estimation, as they depend strongly on the adopted selection criteria as well as on the properties of the systems (e.g., separation, luminosity), but a comparison can still be instructive.

\cite{Barrows2023} reported a dual AGN fractions between 0.2\% and 1\%, depending on bolometric luminosity and separation, within the redshift range 0.08 $< z <$ 0.58 and projected separations ranging between 10 and 70 kpc. These values correspond to angular separations between $\sim\,$~\ang{;;1} and $\sim\,$~\ang{;;0.6}, i.e. an order of magnitude larger than the physical scales probed in the present work.
\citet{Li2024} estimated that the pair fraction increases with redshift, starting at approximately 4.5\% at $z \sim 0.5$ and reaching around 22.9\% at $z \sim 4.5$. This fraction drops to below 1\% when considering only objects with projected separations smaller than ~\ang{;;0.8} as in this work. They found the distribution of projected separations to be relatively flat, but did not identify any pairs with separations smaller than $\sim\,$2~kpc.

Recently \cite{Silverman2020} found a dual QSO fraction of (0.26 $\pm$ 0.18)\% from $z = 3$ to $z = 1.5$ with separations of $\ang{;;0.6}$ and ~\ang{;;4}.  
\cite{Yu2011} found that 0.02\%–0.06\% of AGN are dual AGN with double-peaked narrow line features at redshifts of $z$ $\sim\,$ 0.5--1.2.
\cite{Rosario2011} estimated a dual fraction between 0.32\% and 0.5\%, similar to \cite{Fu2011}. \citet{Koss2012} estimated that, among AGN in the \textit{Swift} BAT sample, the fraction of dual AGN at projected separations of less than 15~kpc is approximately 5\% (8 out of 167 systems). In contrast, the corresponding dual AGN fraction among Seyfert galaxies in the SDSS is significantly lower, at only 0.25\%. These estimates are all broadly comparable with our results.

\subsection{Other architectures}
We explored the use of Domain-Adversarial Neural Networks (DANN) for domain adaptation, a technique designed to help models generalize when the target domain data (i.e. the real data) is entirely unlabeled, as in our case \citep{Ganin2015}. The goal of DANN is to learn a mapping between domains such that the features extracted by the network are both discriminative for the task and invariant to domain shifts. This means that the features should have similar distributions in both the source (simulations) and target (data) domains, allowing the model to generalise well to the target domain \citep{Ajakan2014,Huertas-Company2024,Belfiore2025,Ginolfi2025}.
However, our attempt did not provide any improvement on the real data, though the network was able to generalize the features of data and simulations. Minimising the discriminator, we found no improvement in terms of reducing FPs labelled with the visual inspection. 

We also used one of the most popular architectures, Resnet18 \citep{He2016}, by adding the last layers for the prediction of system properties as described in Sect.~\ref{Sample}. Although it achieved a training set accuracy of 99\%, which was not achieved by the other networks, its performance on the test set and on the real data was worse than the network we used, producing an order of magnitude more FPs compared to our CNN, probably due to overfitting. 

We also tried combining different simulations, increasing the size of the training set, but the performance remained similar.

\section{Conclusions}
In this work, we took advantage of the \Euclid data release Q1, which, thanks to its wide area, high sensitivity and resolution,  provides a unique opportunity to identify close AGN pairs. We developed a CNN trained on realistic simulations to search for dual or lensed AGN at sub-arcsec separations (<\,~\ang{;;0.8}). The CNN performs significantly better than classical detection methods, especially in the regime of small separations, and maintains a high completeness while finding a limited number of FPs. Applying our model to a sample of  $\sim\,$6000 QSOs, we identified a first set of promising dual/lensed AGN candidates. Once we estimated the number of foreground contaminants and expected FPs from the CNN, and identified two possible lens candidates, we found a dual fraction of $\sim\,$0.25\%, with an overdensity of these systems at low separations (below \ang{;;0.45}) where contamination of foreground objects is lower compared to high separations. For sources at larger separations ($>$~\ang{;;0.45}) we performed a photometric analysis to study colours relative to \cite{Q1-SP027} diagrams and found that three objects could be compatible with being dual AGN, while most of the systems are most likely foreground contaminants. 
While our CNN demonstrates promising performance for future \Euclid release, further progress requires refining the simulated training set, such as the exploitation of all four \Euclid bands into the machine learning framework. This will not only improve the prediction of the network by providing more realistic examples, but also allow the CNN to estimate the probability that a secondary object is a foreground contaminant, such as a star, or another AGN. Larger samples and the full depth and coverage of future \Euclid releases, such as DR1, should be used to test the scalability of the CNN. The main bottleneck remains in the visual inspection stage, which will not be feasible for future releases. However, because our CNN has been designed to maximise the number of TPs compared to FPs, and given that the expected fraction of dual AGN is relatively small \citep{Rosas-Guevara2019,Volonteri2022}, the number of candidates should be manageable for visual inspection. Spectroscopic follow-up and higher-resolution imaging will be critical to confirm the nature of candidates, especially at sub-arcsec separations. Furthermore, future \Euclid releases, complemented by data from \textit{JWST}, LSST, and \textit{Roman}, will be key to achieving a statistical census of dual AGN and constraining models of SMBH growth and galaxy evolution.  

\begin{acknowledgements}
\AckEC  
\AckQone

This publication was produced while attending the PhD program in PhD in Space Science and Technology at the University of Trento, Cycle XXXIX, with the support of a scholarship financed by the Ministerial Decree no. 118 of 2nd march 2023, based on the NRRP -- funded by the European Union -- NextGenerationEU -- Mission 4 "Education and Research", Component 1 "Enhancement of the offer of educational services: from nurseries to universities” -- Investment 4.1 “Extension of the number of research doctorates and innovative doctorates for public administration and cultural heritage”

We acknowledge financial contribution from INAF Large Grant “Dual and binary supermassive black holes in the multi-messenger era: from galaxy mergers to gravitational waves” (Bando Ricerca Fondamentale INAF 2022), from the INAF project "VLT-MOONS" CRAM 1.05.03.07, from the French National Research Agency (grant ANR-21-CE31-0026, project MBH\_waves), from the INAF Large Grant 2022 "The metal circle: a new sharp view of the baryon cycle up to Cosmic Dawn with the latest generation IFU facilities".
This work has been partially financed by the European Union with the Next Generation EU plan, through PRIN-MUR project "PROMETEUS" (202223XPZM).
A.F acknowledges INAF Mini Grant 2024 "The pc-scale view of HII regions in M33". 
EB acknowledges funding through the INAF GO grant ``A JWST/MIRI MIRACLE: Mid-IR Activity of Circumnuclear Line Emission'' and  the ``Ricerca Fondamentale 2024'' program (mini-grant 1.05.24.07.01).


This research used the facilities of the Italian Center for Astronomical Archive (IA2) operated by INAF at the Astronomical Observatory of Trieste.

FR acknowledges the support from the INAF Large Grant "AGN \& \Euclid: a close entanglement" Ob. Fu. 01.05.23.01.14.
\end{acknowledgements}

\bibliographystyle{aa}
\bibliography{references,Euclid,Q1} 

\appendix
\newpage
\section{Comparison between data and simulations}\label{sec:comparison}

\begin{figure}[h!]
\centering
    \includegraphics[width = 0.47 \textwidth]{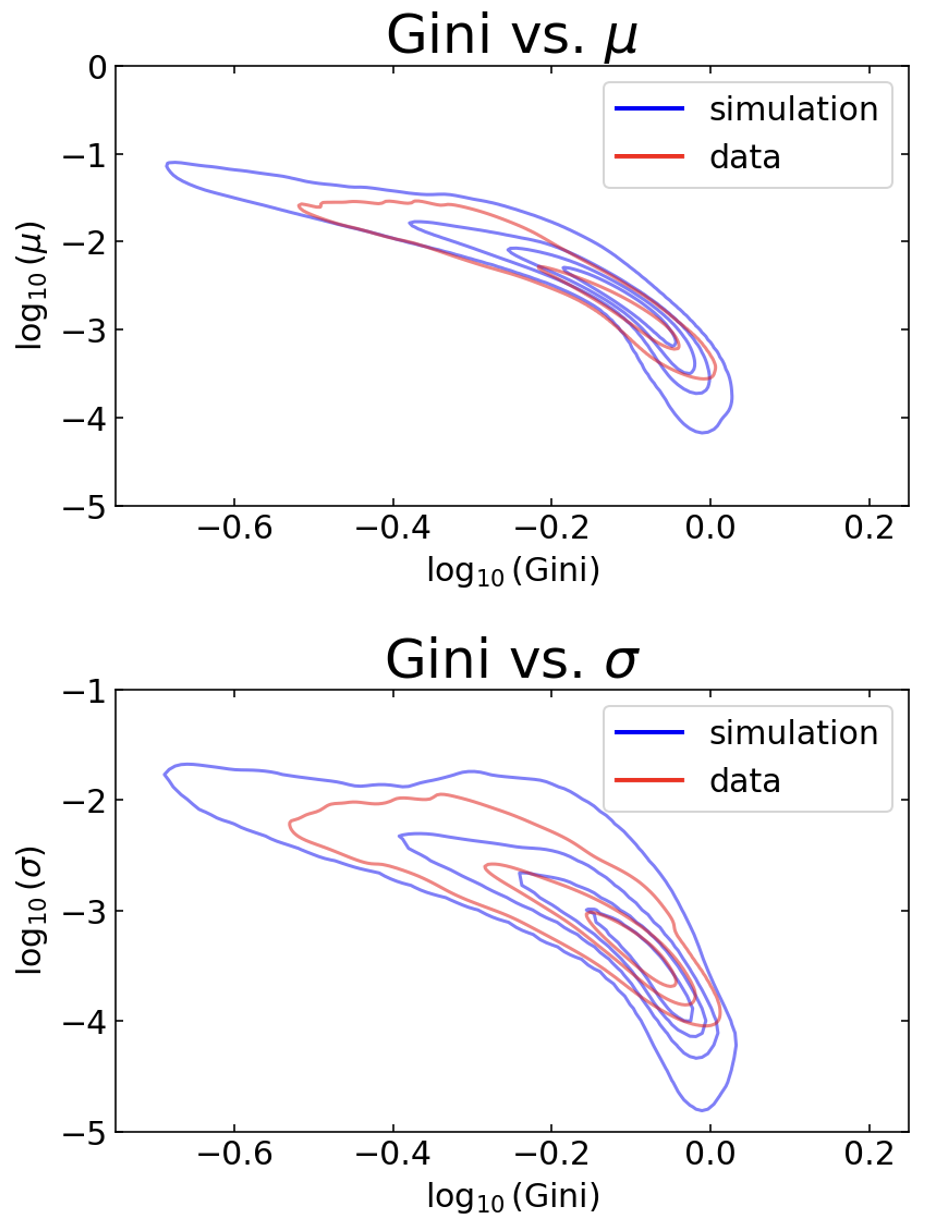}
    \caption{Comparison of the statistical properties of the simulations (blue) and the data (red). 
    The upper panel shows the Gini coefficient vs. mean of the signal in logarithm.  
    The lower panel shows the Gini coefficient vs. standard deviation of the signal in logarithm. 
In both panels, the contours represent kernel density estimates of the respective distributions with 5 levels. }
    \label{fig:Gini}
\end{figure}

We compare the statistical properties of our simulated AGN with those measured on observational data using the mean signal level, the standard deviation, and the Gini coefficient. The Gini coefficient is defined as

\begin{equation}
G = \frac{\sum_{i=1}^{n} \sum_{j=1}^{n} |x_i - x_j|}{2n^2 \bar{x}},
\end{equation}
where \( x_i \) is the flux in pixel \( i \), \( n \) is the total number of pixels, and \( \bar{x} \) is the mean flux. It defines the flux distribution in an image; higher values indicate that the flux is concentrated in the center (for instance, when the image is dominated by an unresolved AGN), while lower values correspond to AGN with more diffuse host galaxy. From Fig.~\ref{fig:Gini} we see that the simulations reproduce the observed distributions well, with both samples occupying the same region of parameter space.

\section{CNN on foreground contamination}\label{appendix:foreground}
\begin{figure*}
    \centering
    \includegraphics[width=1\linewidth]{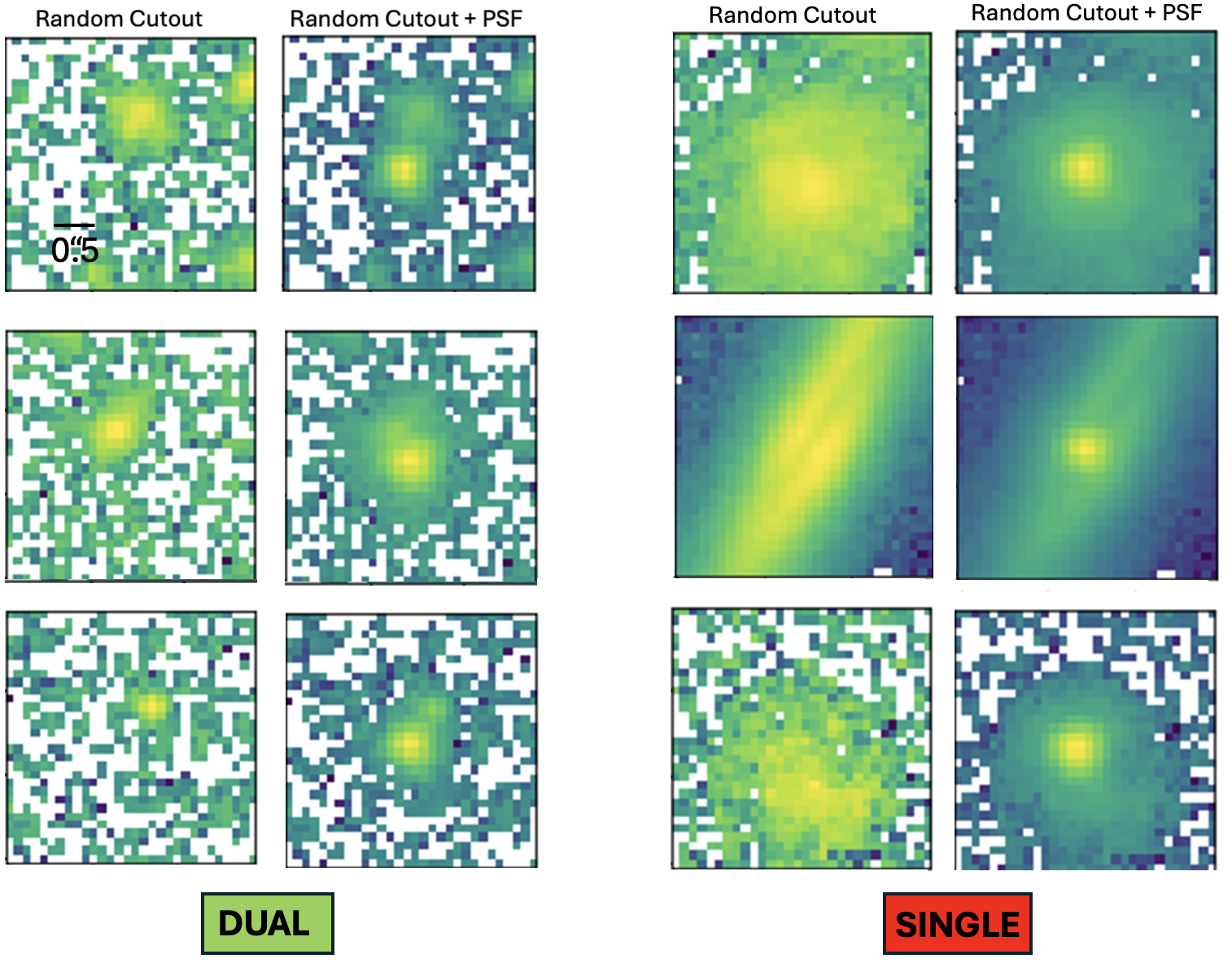}
    \caption{Application of the CNN to the foreground contaminations. 
    The left panel shows examples predicted as dual systems; the right panel shows examples predicted as single systems. 
    In both panels, the first column represents random cutouts and the second column the random cutouts + PSF.}
    \label{fig:foreground_cont}
\end{figure*}

In Sect.~\ref{sec:foreground_cont} we gave an estimate of foreground contaminants. In order to compare the number of dual objects found in Q1 and the number of possible contaminants, it is necessary to use the same detection algorithm, in this case CNN. Therefore, once we added a PSF to the random cutouts, it was possible to study how the network behaves in the case of extended contaminants sources. In Fig.~\ref{fig:foreground_cont}, we show some examples of the behavior of the network recognising when the contamination is point-like (left panel, dual) or extended and not considered as a contaminant (right panel, single).

\section{Lens candidates}\label{appendix:lenscandidates}
\begin{figure}
    \centering
    \includegraphics[width=1\linewidth]{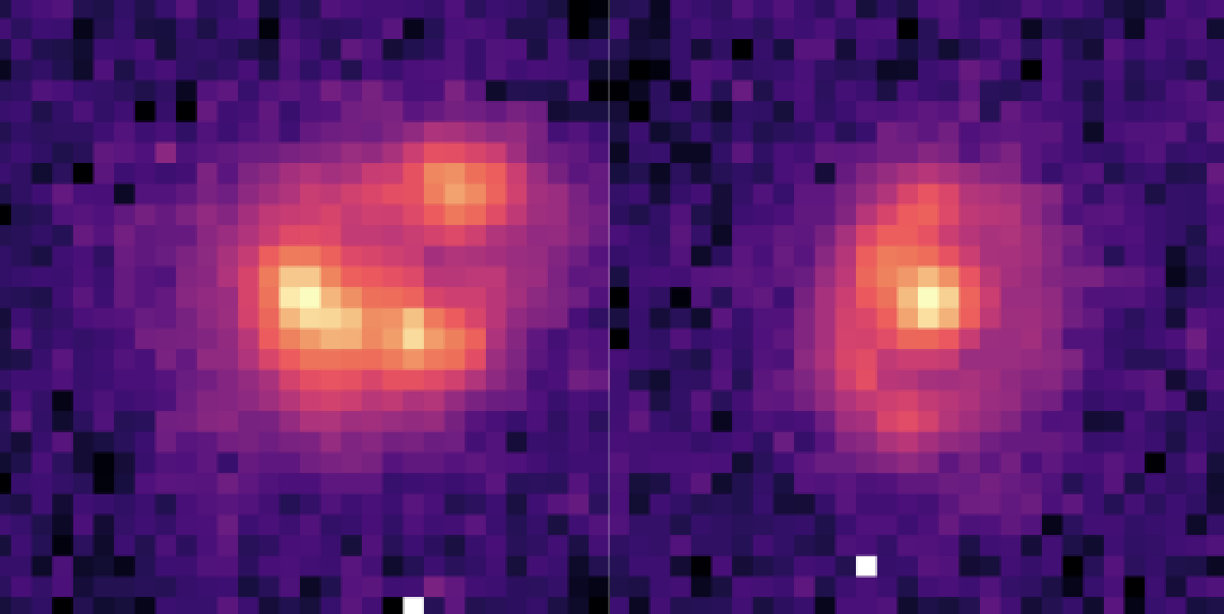}
    \caption{VIS images of objects predicted as dual by the CNN showing lens-like configurations.}
    \label{fig:possible_lens}
\end{figure}

The network classified both the objects in Fig.~\ref{fig:possible_lens} as dual candidates. In the visual inspection, we identified features consistent with lensing and we classified both as class 3. The first system may represent a quadruply lensed configuration, while the second exhibits an arc-like structure that could also be interpreted as a ring galaxy, similar to the case shown in Fig.~\ref{fig:high_res_facilities}(c).

\label{lastpage}
\end{document}